\documentclass[12pt]{article}                                   % -*- latex -*-

\usepackage{times}
\usepackage[hyperfootnotes=false]{hyperref} % url package comes with this
\usepackage{type1cm} % We are using scalable fonts. Really.
\usepackage{amsmath}

\usepackage{texlib/ct}
\usepackage{texlib/code}
\usepackage{texlib/headings}	% Tighter section titles

\hypersetup{linkcolor=black,urlcolor=black,citecolor=black,pdfborder=0 0 0}

\input{decls}

\newcommand{\Section}[1]{\section*{#1}}

\newcommand{\Bibliography}[1]{\clearpage\bibliography{#1}}

\title{An Introduction to \\ Rank-polymorphic Programming \\
       in Remora \\
       (Draft)
       }
\author{Olin Shivers, Justin Slepak, Panagiotis Manolios \\
        Northeastern University}

\newenvironment{oldtabular}{\begin{tabular}}{\end{tabular}}

\begin{document}
\maketitle
% -*- latex -*-

\begin{abstract}
%Remora is a statically typed, higher-order functional language
%based on the rank-polymorphic model of computation 
%invented by Kenneth Iverson for his language APL and its successor J\@.
%We describe both Remora and
%the general, underlying rank-polymorphic computational model,
%introducing the concepts of the language by means of a sequence
%of tutorial examples.

Remora is a higher-order, rank-polymorphic array-processing
programming language, 
in the same general class of languages as APL and J\@.
It is intended for writing programs to be executed on parallel hardware.

We provide an example-driven introduction to the language and 
its general computational model, 
originally developed by Iverson for APL\@.
We begin with Dynamic Remora, 
a variant of the language with a dynamic type system
(as in Scheme or Lisp), 
to introduce the fundamental computational mechanisms of the language, 
then shift to Explicitly Typed Remora, 
a variant of the language with a static, dependent type system 
that permits the shape of the arrays being computed 
to be captured at compile time.

This article can be considered an introduction to 
the general topic of the rank-polymorphic array-processing computational model,
above and beyond the specific details of the Remora language.
A reader generally interested in the topic of the computational model
that serves as the foundation for this entire class of languages
should find the tutorial informative.

We do not address the details of type \emph{inference} in Remora, 
that is, the assignment of explicit types to programs 
written without such annotations; this is ongoing research~\cite{Slepak:PhD}.
\end{abstract}

%%%%%%%%%%%%%%%%%%%%%%%%%%%%%%%%%%%%%%%%%%%%%%%%%%%%%%%%%%%%%%%%%%%%%%%%%%%%%%%
\Section{Introduction}
The rank-polymorphic programming model was developed by Kenneth Iverson, 
first for his array-programming language APL~\cite{APL}, 
and then later refined for its successor J~\cite{J}.
We have subsequently designed a statically typed, higher-order
functional language, Remora, based on the same computational model.
The primary aim of this article is to introduce the reader,
in a gradual, gentle way,
to programming in Remora.
We'll also discuss the pragmatics of Remora programming,
in particular, how the parallel semantics of the language permits
efficient code to be written for execution on parallel hardware.
Finally, the design of Remora can serve as an introduction to the
general model of computation that it embodies,
which may make it easier to learn other languages in the class.
A formal semantics for Remora, 
both its dynamic semantics and static type system,
is presented in other work~\cite{Slepak+:ESOP14,
                        Slepak+:semantics-rank-poly-preprint}.

Rank-polymorphic languages are known for not requiring explicit
iteration or recursion constructs.
Instead, the ``iteration space'' of a program is made real, or ``reified,''
in the shape of its aggregate data structures:
when a function that processes an individual element of this space
is applied to such a data structure,
it is automatically lifted by a general polymorphic mechanism to
apply across all of the elements of the aggregate.

In this tutorial, we'll look at the three core mechanisms that
exist in Remora that work together to constitute its
control story\footnote{Other languages, such as Python, Matlab, or R,
have \emph{ad hoc} mechanisms that permit programmers to do some of
the same things, but without the same generality and design integrity
of languages centrally based on Iverson's computational model.}:
\begin{itemize}
\item Frame polymorphism
\item Principal-frame cell replication
\item Reranking
\end{itemize}
The interplay of these mechanisms permits sophisticated Remora
(or APL, or J) programmers to write programs that are startlingly succinct.

Additionally, we will explore Remora's static type system,
which permits the language to describe, at compile time,
the dimensions of the arrays computed by Remora programs.
We'll begin this tutorial by avoiding any mention of static types;
once the dynamic mechanisms of the language are understood,
we'll move on to the issue of how to capture the shapes of arrays
with a static semantics,
in ways that respect the above three mechanisms 

\Section{Everything is an array}

In rank-polymorphic languages such as Remora,
\emph{all values are arrays}.
That is, every Remora expression evaluates to an array.
An array is a collection of data arranged in a hyper-rectangle of some
given dimensionality.
Every array comes with its constituent elements, and a shape.
Array elements come from a separate universe of \emph{atoms};
typical atoms are numbers, characters, booleans and functions.
Permitting arrays of functions means that Remora is a higher-order
functional language.

For example, consider a matrix that has two rows and three columns of integers
\[
\left[\begin{array}{ccc} 7 & 1 & 2 \\ 2 & 0 & 5\end{array}\right]
\]
We say that this matrix has \emph{rank} 2---that is,
it has two dimensions or axes of indexing---and
\emph{shape} $[2,3]$.
The shape of an array is a sequence (or, equivalently, list or vector)
giving its dimensions.

As another example, suppose we have collected rainfall data showing
the monthly rainfall for twelve months of the year,
across fifteen years of data collection,
for all fifty states of the USA\@.
We could collect this data as a numeric array \ex{RF}
of rank 3 and shape $[50,15,12]$.

In principle, we could pull out the rainfall for April (month 3) of
year 6 for the state of Georgia (state \#9)
by indexing into the array with the appropriate
indices: \ex{RF[9,6,3]}.
But well-written programs in rank-polymorphic languages do not operate
on individual elements of arrays;
as we'll see, programs operate on entire arrays.
So indexing is, in fact, something upon which we frown.\footnote{
  It's not impossible to do:
  as we'll see in a later section (page~\pageref{sec:indexing}),
  Remora does have an indexing operator that works in a data-parallel way.
  It's best to think of indexing in Remora as a
  fairly heavyweight communications operation that permits programmers
  to shuffle entire collections of data,
  not as a means of accessing array elements one-at-a-time.}

The rank of a scalar array is 0 and its shape is the empty vector $[\,]$.
Note that:
\begin{itemize}
\item The rank of an array is also the length of its shape,
      which is maintained in the case of scalar values.
\item Multiplying together the numbers in the shape of an array
      tells us how many atoms the array contains. For example,
      the shape of our rainfall-data array is $[50,15,12]$, so
      the array contains $50 \times 15 \times 12 = 9000$ elements.

      As a boundary case, consider the scalar array whose only
      element is the number 17. The shape vector for a scalar
      array is the empty vector $[\,]$; multiplying all the elements
      of the empty vector together produces $1$, which is, indeed,
      the number of elements contained by a scalar array.
\end{itemize}

In Remora, a language with a \textsc{Lisp}-like s-expression syntax,
the primitive notation for writing a literal array is the \ex{array} form,
that gives the shape of the array followed by its elements listed
in row-major order.
So our two example arrays, above, along with the scalar 17,
could be written in Remora as the constant expressions
\begin{code}
(array [2 3] 7 1 2 2 0 5) ; Our 2x3 example matrix
(array [50 15 12]         ; Rainfall data
       8 14 10 10 \ldots)    ;   9,000 elements here
(array [] 17)             ; The scalar value seventeen\end{code}
Note that Remora's basic s-expression syntax uses square brackets
as well as parentheses; these are notationally distinct.
Note, also, the \textsc{Lisp} comment syntax:
all text from a semicolon to the end of a line is ignored.

{\newcommand{\dpr}[1]{\ensuremath{{d^{\prime}}_{#1}}}

Array-producing expressions can be assembled into larger arrays with
the \ex{frame} form:
\begin{code}
(frame [\vi{d}{1} \ldots] \vi{e}{1} \ldots)\end{code}
The first subform of a \ex{frame} expression is a shape or list of dimensions
\texttt{[$d_1$ {\ldots} $d_n$]}.
This is followed by as many expressions as the product of the $d_i$;
these must all produce arrays of
identical shape \ex{[\dpr{1} {\ldots} \dpr{m}]}.
Once these expressions have been evaluated,
their result arrays are assembled together to produce a final array
of rank $n+m$ and
shape \ex{[$d_1$ {\ldots} $d_n$ $\dpr{1}$ {\ldots} $\dpr{m}$]}.

}

For example, the following code defines \ex{v} to be a 3-element vector,
and \ex{m} to be a two-row, three-column matrix whose two rows are
each identical to \ex{v}:
\begin{code}
    (define v (array [3] 8 1 7)) ; Shape [3]
    (define m (frame [2] v v))   ; Shape [2 3]\end{code}

Note the distinctions between the \ex{array} and \ex{frame} forms.
The \ex{array} form is for writing down array \emph{constants},
that is, \emph{literal} arrays; its subforms are literal atoms.
The \ex{frame} form causes run-time computation to occur:
  we evaluate the expressions that are its subforms to produce
  arrays that are then ``plugged into'' position in the given
  frame to make a larger, result array.

Now that we've introduced the \ex{array} and \ex{frame} forms,
we'll hide them from view at every turn 
by means of some convenient syntactic sugar:
\begin{itemize}
\item First, whenever an atom (that is, an array \emph{element}) literal $a$
appears in a syntactic context where we expect an expression\footnote{Remember:
all expressions produce arrays.},
it is taken to be a scalar array---that is,
it is treated as shorthand for \ex{(array [] $a$)}.
\item Second, whenever a sequence of expressions occurs surrounded
by square brackets in an expression context,
it is treated as a \ex{frame} form for a vector frame.
That is, the expression \ex{[$e_1$ {\ldots} $e_n$]}
is treated as shorthand for \ex{(frame [$n$] $e_1$ {\ldots} $e_n$)}.
\item Finally, a frame whose component expressions are all array literals
is, itself, collapsed to a single \ex{array} term.
\end{itemize}

Thus we could write the scalar array 17 as expression \ex{17},
and the vector of the first five primes as expression \ex{[2 3 5 7 11]};
our original example array could be written as
\begin{code}
[[7 1 2]    ; A 2x3 matrix
 [2 0 5]]\end{code}
This is \emph{exactly} equivalent to the array-literal expression
\begin{code}
(array [2 3] 7 1 2 2 0 5)\end{code}

Likewise, we could write the 
truth table for $i \text{ xor } j \text{ xor } k$,
using $0$ for false and $1$ for true, as the rank-3 array
\begin{code}
;;; A 2x2x2 array
[[[0 1]   ; i=0 plane / j=0 row
  [1 0]]  ; i=0 plane / j=1 row

 [[1 0]   ; i=1 plane / j=0 row
  [0 1]]] ; i=1 plane / j=1 row\end{code}

%[[[ 9 14 10 10 9 7 6 6 7 7 8 9] ; first year of state 1
%  [10 14 11 10 9 8 7 6 7 8 9 9] ; second year of state 1
%  \ldots]
%
% [[ 7 12  7  7 7 6 5 5 6 7 7 8] ; first year of state 2
%  [ 8 12  7  7 8 7 6 6 7 7 7 7] ; second year of state 2
%  \ldots]
%
% ;; Data for states 3-50 elided.
% \ldots]\end{code}

When using the square-bracket notation, the shape of the array
is determined from the nesting structure of the expression.
It's not allowed for two brother elements in a square-bracket
array expression to have different shapes; they must match.
Thus, the following ``ragged'' matrix is not a legal expression,
as it doesn't have a well-defined shape:
\begin{code}
[[7 1 2]
 [9 5]    ; Illegal -- row too short!
 [2 0 5]] \end{code}
As we'll see later (page~\pageref{sec:boxes}),
there is a mechanism in Remora called a ``box,''
that permits programmers to make ragged arrays, but we'll ignore this for now.

\Section{Functions operate on ``cells'' of input}

In Remora, every function is defined to operate on arguments of
a given rank and produce a result of a given rank;
these are called the \emph{cells} of the function application.
For example, the addition operator \ex{+} operates on two arguments,
each of which is a scalar, that is, of rank 0.
\begin{code}
    (+ 3 4)
\result{7}
    (+ 2 8)
\result{10}\end{code}
In this example, and the examples to come, 
we'll show code and the result expressions it produces,
in an ``interactive'' style, 
as if we were presenting Remora expressions and definitions
to an interpreter:
the input Remora expression will be indented, and
the value produced will displayed, flush left, 
on the following line.

As further examples,
we could have a dot-product function \ex{dot-prod} that
operates on two arguments of rank 1;
or a polynomial evaluation function \ex{poly-eval} that operates
on a vector (rank 1) giving the coefficients of a polynomial,
and a scalar (rank 0) giving the $x$ value where we are
evaluating the polynomial:
\begin{code}
    (dot-product [2 0 1] [1 2 3])
\result{5}
    ;; Evaluate 2 + 0x - 3x^2 at x=1
    (poly-eval [2 0 -3] 1)
\result{-1}\end{code}

The argument ranks of a function are part of its static definition;
    when we define our own functions, we must specify them.
We do this by tagging each parameter to the function with its rank.
So, both \ex{x} and \ex{y} inputs to the \ex{diff-square} function below
are specified as being of rank 0:
\begin{code}
    (define (diff-square [x 0] [y 0])
      (- (* x x)
         (* y y)))

    (diff-square 5 3)
\result{16}\end{code}

\Section{Functions distribute over a frame of cells}
The fundamental, core iteration mechanism of Remora is that
any function defined to take arrays of rank $r$
is automatically lifted by the language so that it can be applied
to arrays of any rank $r' \geq r$.
This is the polymorphism of ``rank polymorphism.''

This lifting is accomplished by viewing an array as
a \emph{frame} of \emph{cells}.
Consider our $2\times 3$ matrix:
\[
\left[\begin{array}{ccc} 7 & 1 & 2 \\ 2 & 0 & 5\end{array}\right]
\]
We can view this array three ways:
\begin{enumerate}
\item as a scalar frame containing a single $2\times 3$ matrix cell;
\item as a vector frame containing two cells that are 3-vectors; or
\item as a $2\times 3$ matrix frame containing six scalar cells which
    are the individual elements of the array.
\end{enumerate}
In general, an array with rank $r$ and shape $s = [d_1,\ldots,d_r]$
can be viewed as a frame of cells in $r+1$ ways, depending on where
one splits the shape into the frame prefix and the cell suffix.
%For example, here are the three frame/cellsplits of the $2\times 3$
%matrix shape, with ``$|$'' marking the division between the frame
%and the cell parts of the shape.
%\begin{inset}
%\marginpar{This is ugly. Tweak this.}
%\begin{tabular}{ll}
%$[\;|\;2,3]$ & scalar frame of a (single) $[2,3]$ matrix cell \\
%$[2\;|\;3]$ & vector frame (of length 2) of vector cells (of length 3) \\
%$[2,3\;|\;]$ & $2\times3$ matrix frame of scalar cells
%\end{tabular}
%\end{inset}

When we apply a function that consumes argument cells of rank $r\,$
to an actual argument array of rank $r' \geq r$,
we divide the input into a frame of cells:
the cells have rank $r$, and the frame has rank $r'-r$.
The shape of each cell is given by
the last $r$ dimensions of the array's shape;
the shape of the frame is the remaining, initial $r'-r$ dimensions
of the array.
That is, the function imposes a view on the argument array:
we view the rank $r'$ array as a collection of rank $r$ argument cells.

The function is then applied, in parallel, to each argument cell;
the results of all these independent applications
(which must all have the same shape)
are then collected into the frame to produce the final result.

For example, suppose we have a function \ex{vmag} that takes a vector
(that is, an array of rank 1) and returns its Euclidean length or magnitude:
\begin{code}
    (define (vmag [v 1]) \ldots)
    (vmag [3 4])
\result{5}
    (vmag [1 2 2])
\result{3}\end{code}

Note that the \ex{v} parameter to \ex{vmag} is defined to take arguments
of rank 1.
If we apply \ex{vmag} to a matrix, it is applied independently to each
row of the matrix. That is, \ex{vmag} views the matrix as a vector
frame of vector cells.
All the scalar results of these \ex{vmag} applications are collected into
 the original argument's vector frame:
\begin{code}
    (vmag [[1 2 2]
           [2 3 6]])
\result{[3 7]}\end{code}

Likewise, if we applied \ex{vmag} to a six-dimensional array,
it would be treated as a five-dimensional frame of vector cells;
each such cell would have its length computed,
and we would collect these scalar answers into the frame
to produce a rank-5 array result.

\Section{Multiple arguments and frame agreement}

The frame-distribution mechanism of function application applies
just as well when a function has multiple arguments.
For example, consider our polynomial-evaluation function, \ex{poly-eval},
that takes a vector of coefficients and a scalar value at which we
wish to evaluate the polynomial.
Suppose we apply this function to a matrix and a vector
\begin{code}
    (poly-eval [[2  0 -3]       ; two polynomials
                [5 -1  1]]
               [-2 1])          ; two x values
\result{[-10 5]}\end{code}
The coefficient matrix has shape $[2,3]$,
and the vector of $x$ values has shape $[2]$.
Since \ex{poly-eval} operates on vectors for its first argument,
it views the matrix as a vector frame (shape $[2]$)
of vector cells (shape $[3]$);
likewise, it views its vector of $x$ values as
a vector frame (shape $[2]$) of scalar cells (shape $[\,]$).
Note that once we've pulled off the cell-shape suffixes from the shapes
of each argument, we are left with identical frame shapes: $[2]$.
This is called ``frame agreement,'' which means we have a consistent
frame across which to distribute the individual function applications.
Thus, we evaluate the polynomial $2 + 0x - 3x^2$ at $x = -2$,
and $5 - x + x^2$ at $x=1$, collecting the results into the vector frame
and producing final answer \ex{[-10 5]}.
Table~\ref{table:poly-eval-frame} shows how the cell requirements of
\ex{poly-eval} guide the distribution of the function call over the
collection of inputs it is given.
\begin{table}
\begin{center}
\begin{tabular}{l@{\quad}l@{\quad}l@{\quad}l}
Argument                & Argument shape        & Frame shape   & Cell shape \\
\hline
\#1 (coefficients)      & $[2,3]$               & $[2]$         & $[3]$ \\
\#2 (x)                 & $[2]$                 & $[2]$         & $[\,]$
\end{tabular}
\end{center}
\caption{The \ex{poly-eval} function imposes a frame/cell view on each
of its arguments. After removing the cell-shape suffix from the shape
of each argument array, the remaining shape prefix is the argument's frame.
Both arguments in our example have the same frame, which means we can lift the
function application across the two matching frames;
this frame then provides the outer structure into which we collect the
results of the distinct invocations of \ex{poly-eval}.
}
\label{table:poly-eval-frame}
\end{table}

Remora's lifting mechanism essentially distributes the \ex{poly-eval}
function across the two matching frames of arguments in a pointwise fashion,
producing one function invocation for every cell of the frame.
It is as if we had written
\begin{code}
    [ (poly-eval [2  0 -3] -2)
      (poly-eval [5 -1  1]  1) ]
\result{[-10 5]}\end{code}

\Section{The principal frame and cell replication}

Remora's frame-based distribution mechanism is more general than
simply requiring the frames of all argument arrays to match.
The full rule is driven by the notion of an application's ``principal frame.''
In a given function application,
the argument frame with the \emph{longest shape} is
considered the principal frame;
for the function application to be well-formed,
the frame shapes of all other arguments
must be a \emph{prefix} of the principal frame's.
When distributing an argument's cells 
across the cell-wise invocations of the function,
if an argument's frame has a shape shorter than
(a proper prefix of) the principal frame's,
then the array is replicated into the missing dimensions to provide
enough cells for the full frame of function applications.

For example, suppose we add a vector of 2 numbers to a $2 \times 3$ matrix:
\begin{code}
    (+ [10 20] [[8 1 3]
                [5 0 9]])
\result{[[18 11 13]
 [25 20 29]]}\end{code}
The addition operator adds two scalars to produce a scalar result.
Since the two frames have shapes $[2]$ and $[2, 3]$, 
the principal frame's shape is $[2, 3]$.
The first argument's frame gets replicated from shape $[2]$
to shape $[2, 3]$.
The way to think of this replication is that
when we select a cell from this argument, for frame element $[i,j]$,
we simply \emph{drop} any suffix of the index not needed
to index into the argument's actual frame---in this case, the column index $j$.
This means that we match every column of the right argument's first row with
the first element of the left vector: $10+8$, $10+1$, and $10+3$;
likewise, we match the items of the right argument's second row with
the second element of the left vector: $20+5$, $20+0$, $20+9$.
Thus, we get one function application for each element of the principal frame,
where the results are collected,
producing an answer which is a $2 \times 3$ frame of scalar cells.

In short, the frame-agreement rule of Remora means that
when we add a vector to a matrix,
we add the first element of the vector to the first row of the matrix,
the second element of the vector to the second row of the matrix,
and so forth.
(What if we want to add the first element of the vector
to the first \emph{column} of the matrix, and so forth?
We'll come to this later.)

Likewise, if we add a matrix $M$ to a three-dimensional array $A$,
then we add element $M[i,j]$ to each element of plane $i$, row $j$ of
A;
that is, we add each scalar cell $M[i,j]$ to each scalar cell $A[i,j,k]$.

Given this rule, adding a scalar $s$ to any array simply adds $s$
to each element of the array:
\begin{code}
    (+ 10 [7 1 4])
\result{[17 11 14]}
    (+ [7 1 4] 10)
\result{[17 11 14]}\end{code}

If we wish to evaluate a collection of polynomials at a single $x$,
we simply apply the \ex{poly-eval} function to the collection and the $x$ value.
Whereas, if we wish to evaluate a single polynomial
at a collection of $x$ values, we apply the function to the polynomial
and the collection of $x$ values:
\begin{code}
    ;;; Evaluate two polynomials at the same x.
    (poly-eval [[2  0 -3]        ; 2 + 0x - 3x^2
                [5 -1  1]]       ; 5 -  x +  x^2
               -1)               ; x = -1
\result{[-1 7]}
    ;;; Evaluate 2 + 0x - 3x^2 at four values of x.
    (poly-eval [2 0 -3] [[0 1]
                         [2 3]])
\result{[[  2  -1]
 [-10 -25]]}\end{code}
The second example above shows that, in Remora,
the shape of the output collection is determined by
the shape of the input collection:
When we pump a \emph{matrix} of inputs through the same polynomial,
we get a \emph{matrix} of outputs,
since the principal frame comes from the matrix argument.

In none of these cases
do we need to write a loop, or index into a collection of data;
this is managed for us by the rank-polymorphic lifting of the
\ex{poly-eval} function across its arguments.

It's worth noting that this replication mechanism only replicates
\emph{entire cells}.
It does not replicate arrays that are smaller than one entire cell's
worth  of data up into a complete argument cell.
Thus, it will not, for example, permit us to apply our vector
dot-product function to a scalar, such as \ex{0}.
In short, an actual argument must always provide enough data to
a function to comprise at least one complete argument.

\Section{Frame-replication even applies to the function position}

Because Remora is a higher-order functional language, 
we can write a general expression in the function position of 
a function-application expression;
because expressions in Remora evaluate to arrays,
this means that the function position of an application can be an
\emph{array} of functions.
For example,
\ex{+} is a variable whose value in the top-level environment
is a \emph{scalar array}
whose single element is the addition function---as described earlier,
functions in Remora are atoms, that is, array elements,
just as numbers, booleans, and characters are.

Thus, when we evaluate the expression that is the function position
of a function application, we get an array (of functions),
and \emph{this array participates in the determination of the principal frame
for the application, and is subject to frame replication
just as the argument arrays are.}
The ability to apply an array of different functions to an argument
gives Remora a MIMD-style capability to its parallel semantics.

The function position takes scalar cells,
which means that in the common case,
when we apply a scalar function array
(such as the \ex{+} array) to a pair of argument arrays,
it is replicated across all the applications.

But we can use non-scalar arrays of functions, as well.
Here, we apply a matrix of functions to the single value $9$,
collecting the results into the matrix principal frame:
\begin{code}
    (define m [[square  square-root] ; M is a 2x2 array
               [add1    sub1]])      ; of functions

    (m 9) ; Apply all the functions to nine.
\result{[[81 3]
 [10 8]]}\end{code}

\Section{Some functions take the entire argument as cell}

The frame-of-cells story in Remora has a useful corner case:
it is possible to specify that a particular parameter to a function
takes its entire argument as a single cell.
For example, the \ex{append} function takes two arrays
and appends them along their initial dimension.
Appending two matrices appends the rows of the second matrix
after the rows of the first matrix.
So appending a $3 \times 5$ array and a $7 \times 5$ array
produces a $10 \times 5$ result.
Likewise, appending two three-dimensional arrays appends the
planes of the second array after the planes of the first array:
appending a $3 \times 2 \times 5$ array and a $4 \times 2 \times 5$ array
produces a $7 \times 2 \times 5$ result.
\begin{code}
    (define m1 [[0  1]
                [2  3]])\pagebreak[0]

    (define m2 [[10 20]
                [30 40]])
\pagebreak[0]
    ;;; Append two 2x2 matrices; result is 4x2.
    (append m1 m2)
\pagebreak[0]
\result{[[0  1]
 [2  3]
 [10 20]
 [30 40]]}\end{code}

The append function is defined so that it consumes both of its
arguments, of any rank, as a single cell; thus its frame is a scalar.
When we define our own functions,
we declare this by tagging a parameter with the special keyword \ex{all}
instead of a natural number for its cell rank:
\begin{code}
    (define (append [a all] [b all]) \ldots)\end{code}
One way to view such a parameter is that we fix,
not the \emph{cell} rank of the parameter, but its \emph{frame} rank:
such a parameter has a scalar frame.

What if we want to append along a different axis of an array?
For example, instead of appending the two previous matrices one above the other,
suppose we wanted to append them side-by-side, producing:
\begin{code}
\result{[[0  1 10 20]
 [2  3 30 40]]}\end{code}
We'll see how to do this in a following section.

\Section{Iteration with reduce, scan, fold and trace}

An important operator with scalar frame rank is the higher-order \ex{reduce}
function,
which maps an associative binary function \ex{+} 
over the initial dimension of an array.
That is, \ex{reduce} takes an array of non-zero rank $r$
and shape $[d_1, \ldots, d_r]$, 
and considers it a collection of $d_1$ arrays of rank $r-1$ and
shape $[d_2, \ldots, d_r]$.
This collection of arrays $[a_1, \ldots, a_{d_1}]$ is combined
together using some provided function \ex{+},
producing the result
\begin{code}\cdmath
(+ $a_1$ (+ $a_2$ (+ $\ldots$ $a_{d_1}$)))\end{code}
Since \ex{reduce} requires the combining function \ex{+} to be associative,
we can think of the function as operating on all the items at once
\begin{code}\cdmath
(+ $a_1$ $a_2$ $\ldots$ $a_{d_1}$)\end{code}

For example, if we wish to sum the elements of a vector,
we reduce it with \ex{+}:
\begin{code}
    (reduce + [1 4 9 16]) ; Sum of first four squares
\result{30}\end{code}
Similarly, if we wish to multiply the elements together,
we reduce the same vector with \ex{*}:
\begin{code}
    (reduce * [1 4 9 16]) ; Product of first four squares
\result{576}\end{code}
The definition of \ex{reduce} does not specify in what order these
numbers will be multiplied.
It may be done left-to-right
\begin{code}
(* (* (* 1 4) 9) 16)\end{code}
right-to-left
\begin{code}
(* 1 (* 4 (* 9 16)))\end{code}
or in any other order permitted by the axiom of associativity,
such as
\begin{code}
(* (* 1 4) (* 9 16))\end{code}
%Note that the last order permits the \ex{(* 1 4)} and \ex{(* 9 16)}
%combining operations to execute in parallel
%(which is the entire point of the associativity requirement).

If we reduce a matrix with \ex{+},
we will add the first row to the second row, the third row, and so forth.
So, in effect, we will sum each column:
\begin{code}
    (reduce + [[1     2   3]
               [10   20  30]
               [100 200 300]])
\result{[111 222 333]}\end{code}
What if we want to sum along a different axis of the array?
We'll see how to do this in a following section.

The associativity requirement for the combining operator
restricts its type to $\alpha \times \alpha \tu \alpha$.
Note that the operator gets automatically lifted to operate on the subarrays
if it is defined to take cells of smaller rank.
Thus, our \ex{+} operator, which fundamentally operates on scalars,
was lifted to operate on vectors when it was used in the matrix example above.
%Likewise, the initial ``zero'' element
%(which is, in fact, \ex{0} in our example),
%was lifted by \ex{reduce} from its scalar \ex{0} value to the
%required vector \ex{[0 0 0]}.

The \ex{reduce} operator must be applied to an array whose initial
(or ``iteration'') dimension is non-zero.
If we want to sum a matrix \ex{m} that might have zero rows,
\ex{reduce} has no data to feed the \ex{+} operator,
hence no way to produce a result.
If this happens, Remora will report an error.\footnote{A run-time error,
in the case of Dynamic Remora;
a compile-time type error, in the case of Typed Remora.}
In such a case,
the programmer can use an alternate ``reduce with zero'' function,
and explicitly provide a zero value, writing \ex{(reduce/zero + 0 m)}.

Remora also provides a \ex{fold} function that uses a more general
folding operator of type $\alpha \times \beta \tu \beta$,
combining items of type $\alpha$ into a running accumulator of type $\beta$.
For example, we can compute the sum of the magnitudes of a collection of
vectors with
\begin{code}
    (fold (\l{[v 1] [sum 0]} (+ sum (vmag v)))
          0
          [[1 2 2]      ; length 3
           [2 3 6]]))   ; length 7
\result{10}\end{code}
The advantage of using the less general \ex{reduce} is that it permits
the reduction to be performed in a parallel fashion;
\ex{fold} is serial.
So we would be better off expressing the above calculation as:
\begin{code}
(reduce + (vmag [[1 2 2]
                 [2 3 6]]))\end{code}
The individual \ex{vmag} computations can be executed in parallel,
as can the additions of the final summation.
This is unimportant in this example, 
where our matrix represents a collection of only two small vectors, 
but would be significant if we had either a large number of vectors,
or our vectors were extremely long---that is,
if our matrix had a large number of rows or columns (or both).

There is also an \ex{iscan} operator that produces the ``prefix sums''
of an operator applied across the initial dimension of an array:
\begin{code}
    (iscan + [2 10 5]) ; Produce [2, 2+10, 2+10+5]
\result{[2 12 17]}

    (iscan + [[1     2   3]
              [10   20  30]
              [100 200 300]])
\result{[[1     2   3]  ; row 1
 [11   22  33]  ; row 1 + row 2
 [111 222 333]] ; row 1 + row 2 + row 3}\end{code}
The \ex{iscan} function provides an ``interior scan:''
the $i\text{th}$ element of the result includes the
$i\text{th}$ element of the input collection.

An alternative class of scan functions are the ``external'' scans,
which do \emph{not} include the $i\text{th}$ element of the input
collection in element $i$ of the result:
\begin{code}
    (scan/zero + 0 [2 10 5]) ; [0, 2, 2+10, 2+10+5]
\result{[0 2 12 17]}\end{code}
Note that the exterior scan's result is one element longer than
its input:
this is because an external scan includes the initial ``zero'' value
in the result collection.

Remora actually provides three different reduce functions,
eight scans, two folds and two traces.
The trace functions are the ``prefix-sum'' version of the
serial folds;
that is, the trace functions have the same relationship to folding
that the scan functions have to reductions.
The small details differentiating these fifteen functions is
something we won't explore fully, here.
In this tutorial, we'll simply cover these differences as we introduce
needed variants in our example code.

Remora's reduce / scan / fold / trace set of operators provide an important component
of its control story.
The frame/cell lifting mechanism of the language enforces a separation of
computation when we apply a low-rank operator to a high-rank collection of
data.
For example, when we apply the vector-magnitude operator \ex{vmag} to a
three-dimensional array of shape $7 \times 5 \times 3$,
the 35 different applications of \ex{vmag}
all run independently of one another.
This is desireable,
as it permits all the different invocations of \ex{vmag}
to be executed in parallel.
Sometimes, however, we need to perform a computation on a collection of data
that somehow \emph{combines together} the elements of the collection
(a computation with what the scientific-computing community would call
a ``loop-carried dependency'' when expressed in programming languages
that have explicit loops).
In these cases, we use \ex{reduce}, \ex{scan} or one of their serial
brethren---that is their \emph{raison d'\^etre}.

\Section{Remora is a map/reduce architecture: an aside on parallelism}
The use of \ex{reduce} and \ex{scan} characterises a big distinction between
programming in a rank-polymorphic language like Remora and
programming in a serial-array language like \textsc{Fortran}.
In \textsc{Fortran},
we write loops and then hope the compiler can sort out which
computations inside the loop are independent of the iteration order and
can therefore be parallelised,
and which computations have loop-carried dependencies,
and so must be left serialised.
In Remora, the notationally simple way to operate on a collection of data
is simply to apply to the \emph{collection} 
the function that processes a \emph{single} item: \ex{(f collection)},
and this default case is the parallel case.
The \emph{actual semantics} is parallel---we are not just emulating
a serial semantics with a parallel implementation---so the compiler is
licensed to perform all the per-item calculations in parallel;
no heroic analyses are needed to divine this fact.
On the other hand,
the compiler has no difficulty spotting loop-carried dependencies
when they do arise,
because the programmer \emph{explicitly marks} them by writing
down one of the reduce / scan / fold / trace operators.

What is particularly nice about providing reductions and scans as a
central iteration construct is that these operations are a parallelism
slam-dunk.
Suppose we are reducing a collection of $n$ items on a parallel
machine that has $p$ processors.
If we have more processors than data ($n < p$), we can do the operation
in $O(\lg n)$ time.
Since $n$ measures a physical resource,
then here in the physical universe, where most of us reside,
it is entirely reasonable to think of this $O(\lg n)$ time simply as
being able to do a reduction in constant time, for some constant less than 100.
Alternately, in the more common case
that we have more data than processors ($n > p$),
we get good utilisation of the processors:
a speedup of close to $p$, which is all that we can really expect,
given our available hardware resources.

Rank-polymorphic array languages have historically been popular with their
users because the human programmers like the expressiveness and clarity of the
notation, without considering performance.
But it ought to be true that such languages are well-suited to
high-performance implementations on parallel hardware.
(And this is our current research agenda.)

The bottom line is that we should think of Remora as a ``map/reduce''
architecture~\cite{Dean:mapreduce}, 
but one where these parallel operators are notationally
very light\-weight:
every function call comes with a notationally free map,
and reductions are likewise easy to invoke.
Thus, we can program these kinds of computations very easily,
in a context of fine-grained parallelism, 

\Section{Some basic uses of \ex{reduce}}

Here is the definition of the \ex{vmag} function we've
been using in our examples:
\begin{code}
(define (vmag [v 1])
  (square-root (reduce/zero + 0 (square v))))\end{code}
The function consumes vectors,
hence the ``\ex{1}'' rank of its \ex{v} parameter.
We first use the scalar \ex{square} operator to 
produce a vector whose
elements are the squares of the input vector's elements.
Then we sum these elements with a reduction operator,
and take the square-root of the result.
Note that we did this without ever indexing into a vector or writing a loop.

To write the factorial function, we use the primitive \ex{iota} function,
which takes a vector specifying an array shape, and produces
an array of that shape, whose elements are the naturals $0, 1, 2, \ldots$
laid out in row-major order:
\begin{code}
    (iota [5])
\result{[0 1 2 3 4]}
    (iota [2 3])
\result{[[0 1 2]
 [3 4 5]]}
    (+ 1 (iota [5]))
\result{[1 2 3 4 5]}
    (reduce/zero * 1 (+ 1 (iota [5])))  ; 5! = 120
\result{120}
    (define (fact [n 0])\label{example:factorial}
      (reduce/zero * 1 (+ 1 (iota [n]))))
    (fact [0 3 5 10])
\result{[1 6 120 3628800]}\end{code}

\Section{Some simple statistics}

We can average the elements of a vector with this function:
\begin{code}
    (define (mean [xs 1])
      (/ (reduce + xs)
         (length xs)))\end{code}
The \ex{length} function is another function
(like \ex{append} and \ex{reduce})
that consumes its entire argument as its cell;
it returns the size of its argument's initial or leading dimension.
Thus, applying \ex{length} to a $3 \times 5$ array produces $3$.

We can now define variance and covariance using mean:
\begin{code}
    (define (variance [xs 1])
      (mean (square (- xs (mean xs)))))

    (define (covariance [xs 1] [ys 1])
      (mean (* (- xs (mean xs))
               (- ys (mean ys)))))\end{code}
In \ex{variance}, the subtraction operation uses principal-frame replication to
subtract a scalar (the mean of the vector) from each element of the vector.
The scalar \ex{square} function is lifted to apply it
pointwise to all the elements of its vector argument.
Similarly, in \ex{covariance}, the scalar multiply operation \ex{*} is
lifted to pointwise multiply the two argument vectors,
producing a vector result, which is then averaged with \ex{mean}.
All of this is accomplished without needing to write an explicit loop
or array index;
instead of operating on scalar data, the program's operations are applied
to entire collections.

\Section{One-dimensional convolution}
We can convolve a vector of sample data \ex{v} with a weighted window \ex{w}
in three lines of code:\label{example:convolve1}
\begin{code}
    (define (vector-convolve [v 1] [w 1])
      (reduce + (* (rotate v (indices-of w))
                   w)))\end{code}
The key to this function is the lifted \ex{rotate} operation.
The \ex{rotate} function takes an array of rank $r$
and a vector of $r$ rotation amounts, one for each axis of the array.
Thus, if we want to rotate a matrix \ex{m} by 3 rows and 2 columns, we write
\ex{(rotate m [3 2])}.
Here are some examples:
\begin{code}
    (rotate [2 3 4 5 11] [2])
\result{[4 5 11 2 3]}

    (rotate [[2 3 4  5 11]
             [1 4 9 16 25]]
            [1 0]) ; Swap rows; don't rotate cols.
\result{[[1 4 9 16 25]
 [2 3 4  5 11]]}\end{code}

When we rotate some data array \ex{A} using an \emph{array} of rotation vectors,
Remora's rank-polymorphic lifting rules cause it
to rotate \ex{A} by each of the rotation vectors:
\begin{code}
    (rotate [2 3 5 7]      ; Rotate this vector
            [[0] [1] [2]]) ; with 3 different rotations.
\result{[[2 3 5 7]
 [3 5 7 2]
 [5 7 2 3]]}\end{code}
The principal frame of the operation is given by the rotation argument---since
\ex{rotate} consumes its entire first argument as its cell,
the frame for the first argument is always a scalar frame, 
whose shape $[\,]$ is always a prefix of the second argument's frame.
Here, each individual rotation produces a vector result; these three
vectors are collected into the principal frame to produce the final matrix
result.

The \ex{indices-of} function takes an array of any rank $r$,
and produces an array of rank $r+1$.
If we consider the resulting $r+1$-rank array to be
an $r$-rank array of vectors,
each of these vectors is its location in the containing array.
Thus, if \ex{m} is a $2 \times 3$ array, we get a $2 \times 3 \times 2$ result:
\begin{code}
    (indices-of m) ; m is a 2x3 matrix.
\result{[[[0 0]  [0 1]  [0 2]]
 [[1 0]  [1 1]  [1 2]]]}\end{code}
The \ex{indices-of} function is frequently useful,
as it essentially reifies the iteration space of an array.

When we apply \ex{indices-of} to the weights vector \ex{w} of length $n$,
we get an $n \times 1$ matrix \ex{[[0] [1] {\ldots} [\(n-1\)]]}.
When we use this matrix for the rotation argument in the convolution code,
we rotate the vector of sample data \ex{v}
by the rotation distances $0, 1, {\ldots}, n-1$.
This produces the matrix
\[
\left[
\begin{array}{llllll}
v_0     & v_1     & v_2     & \ldots & v_{m-2} & v_{m-1} \\
v_1     & v_2     & v_3     & \ldots & v_{m-1} & v_0     \\
v_2     & v_3     & v_4     & \ldots & v_0     & v_1     \\
\vdots                                                   \\
v_{n-1} & v_n     & v_{n+1} & \ldots & v_{n-3} & v_{n-2} \\
\end{array}\right]\text{.}
\]
That is, the top row is the original sample vector;
the second row is the sample vector rotated once;
the third row is the sample vector rotated twice;
and so forth.
The width of the matrix is the length of the data vector \ex{v};
its height is the length of the weight vector \ex{w};
and each column of the matrix is one sample window's worth of data.
When we multiply this matrix by the weight vector,
the rank-polymorphic lifting rules of Remora multiply the top row
of the matrix by the first weight;
the second row by the second weight;
and so forth.
After this, we simply sum each column of the result,
collapsing the matrix vertically and producing the final convolution vector.

Again, note that we did not have to write explicit loops,
nor did we ever need to use indexing to extract scalar values out of an array,
instead operating on entire aggregates in parallel.

We leave it as an exercise for the interested reader to write
a version of this function that performs a two- or three-dimensional
convolution.\footnote{%
Have fun.
The 2D case requires one additional function call; the 3D case, two more.}

\Section{Reranking gives control of frame/cell factoring}

Remora's fixed frame-replication strategy sometimes doesn't do what we want.
For example,
if we have an $n$-element vector \ex{v} and an $n \times n$ matrix \ex{m},
we can add the first element of \ex{v} to the first row of \ex{m},
the second element of \ex{v} to the second row of \ex{m}, and so forth,
very simply.
The structure of the addition exactly matches the fixed architecture
of Remora's principal-frame replication machinery,
so we only need to write:
\begin{code}
(+ v m) \end{code}

However, suppose we want to add the first element of \ex{v} to the
first \emph{column} of \ex{m}, and so forth?
We manage this by means of $\eta$-expanding the \ex{+} operation,
adjusting the frame/cell split with the cell-rank parameter annotations
on the wrapper $\lambda$ term.
Consider this example:
\begin{code}
(define v [10 100])
(define m [[1 2]
           [3 4]])

((\l{[x 1] [y 1]} (+ x y))
 v
 m)\end{code}
The key point here is that our $\eta$-expanded version of \ex{+} is
\emph{not identical} to our original \ex{+} function.
We've shifted the function parameters from 0-rank cells to 1-rank cells,
as marked by the \ex{1} rank-annotations on the \ex{x} and \ex{y}
formal parameters.
This means that the new function cuts its arguments up into cells
differently from the way that \ex{+} would do it.
The principle frame is given by the two rows of the matrix argument,
so the application is distributed over these rows
and the single cell of the \ex{v} argument is replicated across
this distribution, giving the following sequence of execution steps:
\begin{code}
((\l{[x 1] [y 1]} (+ x y)) [10 100] [[1 2]
                                     [3 4]])
\(\Rightarrow\)
[((\l{[x 1] [y 1]} (+ x y)) [10 100] [1 2])
 ((\l{[x 1] [y 1]} (+ x y)) [10 100] [3 4])]
\(\Rightarrow\)
[(+ [10 100] [1 2])
 (+ [10 100] [3 4])]
\(\Rightarrow\)
[[(+ 10 1) (+ 100 2)]
 [(+ 10 3) (+ 100 4)]]
\(\Rightarrow\)
[[11 102]
 [13 104]]\end{code}
\ldots which is exactly what we wanted.

Manipulating the way Remora's frame-based distribution works with
a re-ranking $\eta$-expansion is a standard idiom when programming
in rank-po\-ly\-mor\-phic languages.
One way to think of this is to bear in mind that function application,
in a rank-polymorphic language, is a more complex mechanism
than in the classic $\lambda$ calculus.
In some sense, every function application comes wrapped in its
own set of nested loops.
As we saw in this example,
when the computation pushes an array argument from the site
of the function application off to the body of the function being called,
the argument is ``cut up'' into a collection of cells, 
and the function application is replicated in parallel across these cells.
All of this implicit loop structure is why we don't have to write
our own, explicit loops.

Programming in a rank-polymorphic language such as APL, J and Remora
involves developing a reflexive understanding of how principal-frame
cell replication causes arguments to be broken up and distributed.
Because programmers frequently tune this mechanism with reranking,
Remora provides a syntactic shorthand for doing so.
We can write the $\eta$-expanded addition term from the above example
with the reranking \verb|~| notation:
\verb|(~(1 1)+ v m)|.
In general, writing
\begin{code}\cdmath\cddollar
~($r_1$ {\ldots} $r_n$)$\v{exp}$\end{code}
desugars to
\begin{code}\cdmath
(\l{[$v_1$ $r_1$] \ldots [$v_n$ $r_n$]} ($\v{exp}$ $v_1$ \ldots $v_n$))\end{code}
for fresh parameters $v_i$.\footnote{
This is almost true: it ignores the possibility that the evaluation of the
function expression \v{exp} might have a side effect of some kind.
In the presence of side effects and a call-by-value semantics,
we must use the safer desugaring
\begin{code}\cdmath
(let ((f \v{exp}))
  (\l{[$v_1$ $r_1$] \ldots [$v_n$ $r_n$]} (f $v_1$ \ldots $v_n$))\end{code}
In practice, the correct, side-effect-safe desugaring almost always
reduces to the more informal one we initially gave,
as the function term being reranked is typically either
a variable or a $\lambda$ term.}
That is, it permits us to specify the cell ranks $r_i$ for the
function's arguments.

Reranking is often useful in the context of the special functions
that consume their entire actual argument as their cell,
such as \ex{append}, \ex{rotate} and the \ex{reduce}/\ex{scan}
family of functions
(which effectively constitute a distinct component of Remora's control story).

For example, recall that \ex{append} assembles its arguments together
along their leading or initial dimension,
so appending two matrices stacks one above the other.
A re-ranked append, however, can assemble two matrices side-by-side:
\begin{code}
    (define m1 [[0 1]
                [2 3]])
    (define m2 [[10 20]
                [30 40]])

    (append m1 m2) ; m1 above m2
[[0  1]
 [2  3]
 [10 20]
 [30 40]]

    (~(1 1)append m1 m2) ; m1 to the left of m2
[[0 1 10 20]
 [2 3 30 40]]\end{code}
The reranked append works by distributing the append across the vector
cells of the two arguments, assembling the results into a vector frame.
After the reranked application, we have the intermediate frame-of-cells result:
\begin{code}
  [(append [0 1] [10 20])
   (append [2 3] [30 40])]\end{code}

%The \ex{rotate} function is similar to \ex{append} in that it rotates
%its first argument along its initial axis, with the rotation amount
%given by its scalar second argument.
%So rotating a matrix rotates its rows vertically;
%each row is moved as an atomic unit.
%(Equivalently, we could say that each column is rotated vertically
%by the same amount as the other columns.)
%\begin{code}
%    (rotate [[0 1 2]    ; Rotate the rows down
%             [3 4 5]    ; by one, and bring the
%             [6 7 8]]   ; bottom row up to the top.
%            1)
%[[6 7 8]
% [0 1 2]
% [3 4 5]]\end{code}
%If we wish to rotate the matrix in a horizontal way, 
%we use reranking to apply the rotation to each row of the
%matrix:
%\begin{code}
%    (~(1 0)rotate [[0 1 2]   ; Rotate the columns right
%                   [3 4 5]   ; by one, and bring the 
%                  [6 7 8]]  ; rightmost col around to the left.
%                 1)\end{code}
%This steps to the intermediate
%\begin{code}
%    [(rotate [0 1 2] 1)
%     (rotate [3 4 5] 1)
%     (rotate [6 7 8] 1)]\end{code}
%The results of the three rotations (each a vector) are collected into
%the vector frame, producing the final result
%\begin{code}
%[[2 0 1]
% [5 3 4]
% [8 6 7]]\end{code}

Reranking is especially useful in the context of the \ex{reduce}/\ex{scan}
family of operators.
The default behavior of \ex{reduce} is to collapse the array argument
along its initial dimension:
\begin{code}
    (reduce + [[0  1   2]     ; Add the first row to 
               [0 10 100]])   ; the second row.
[0 11 102]\end{code}
However, we can sum \emph{across} the matrix by reranking the reduction
to apply the reduction operation independently to each row of the
input:
\begin{code}
    (~(0 1)reduce + [[0  1   2]
                     [0 10 100]])\end{code}
Here, the \ex{reduce} application is distributed across
the vector (rank 1) cells of the data matrix by the ``\ex{1}'' element
of the reranking, giving
\begin{code}
    [(reduce + [0  1   2])
     (reduce + [0 10 100])]\end{code}
which produces final result
\begin{code}
\result{[3 110]}\end{code}
In this manner, we can use reranking to 
control the axes across which we iterate
when performing a reduction on data collections.

\Section{Matrix multiplication}
All of these computational mechanisms---%
rank polymorphism, principal-frame replication, reduction, and reranking---%
come together when
we write the standard  matrix-multiplication function from linear algebra.
We begin by defining a function \ex{v*m} that multiplies
an $n$-element vector $v$ times an $n \times p$ shaped matrix $m$,
producing a $p$-element vector result.
We want the first element of the result
to be the dot product of $v$ with the first column of $m$;
the second element of the result to be the dot product of $v$ with
the second column of $m$, and so forth:
\begin{code}
    (define (v*m [v 1] [m 2]) (reduce/zero + 0 (* v m)))\end{code}

We're done:
to multiply matrix \ex{a} by matrix \ex{b},
we simply apply \ex{v*m} to the two matrices: \ex{(v*m a b)}!
The individual rows of \ex{a} will be taken as the vector cells of the
first argument, and
each one independently multiplied by the entire matrix \ex{b},
which will be taken as a single cell and
replicated across the individual multiplies.

We can package this up with a definition that specifies rank-2
(that is, matrix) inputs as follows:
\begin{code}
    (define (m*m [a 2] [b 2]) (v*m a b))\end{code}
\ldots but note that this is just a reranked \ex{v*m},
so could alternatively define the function this way:
\begin{code}
    (define m*m ~(2 2)v*m)\end{code}

If two lines of code seems overly prolix,
we can write matrix multiply in a single line of code
by pulling all the cell/frame rank manipulation into the
reranking notation:\footnote{
  We like to think of Remora as a bit less terse
  and a bit more readable than classic APL\ldots
  but we'll admit that this definition is pushing
  the boundaries of clarity for non-expert Remora
  programmers.}
\begin{code}
    (define (m*m [a 2] [b 2])
      (~(0 0 2)reduce/zero + 0 (~(1 2)* a b)))\end{code}

\Section{Polynomial evaluation three ways}
At the beginning of this tutorial, we imagined a polynomial-evaluation
function that takes a vector of coefficients,
and an $x$ value at which to evaluate the polynomial.
We can define this function in several ways;
the various definitions illuminate the considerations that apply to
writing efficient, scalable code in Remora.

We begin with a straightforward definition:
\begin{code}
;;; Simple polynomial evaluation
(define (poly-eval [coeffs 1] [x 0])
  (reduce/zero + 0
               (* coeffs
                  (expt x (iota [(length coeffs)])))))\end{code}
The innermost \ex{(iota [(length coeffs)])} term produces a vector
of exponents.
Suppose, for example, that the coefficients vector is length 4,
with elements \ex{[$c_0$ $c_1$ $c_2$ $c_3$]}.
Then this inner expression produces \ex{(iota [4])},
which is the vector \ex{[0 1 2 3]}.
The exponentiation function \ex{expt} raises \ex{x}
to all four of these powers,
producing the vector \ex{[$x^0$ $x^1$ $x^2$ $x^3$]}.
We multiply this vector, point-wise, by the coefficients vector,
and sum the result with a reduction, producing the final answer.

This definition is simple and clear, 
but we do a lot of redundant multiplication when we compute each
power of $x$ independently of the others.
All told, our four-term polynomial example will do $0 + 0 + 1 + 2 = 3$
multiplies to compute the four powers of $x$ that we need---that is,
this code does a quadratic number of multiplies.

For a four-term polynomial, this is not much of a problem,
but if our polynomial has a hundred terms, 
it is a significant waste of computation.
We would rather do the exponentiations for our hundred-term polynomial
using a total of 98 multiplies,
instead of the 4,851 multiplies that this definition performs.

If we compute our polynomial with Horner's rule,
that is, $c_0 + x (c_1 + x(c_2 + x c_3)$,
we'll only require a linear number of multiplications.
This gives us the following definition, 
which directly instantiates Horner's rule using a right-to-left fold
along the coefficients vector:
\begin{code}
;;; Efficient on serial processor
(define (poly-eval [coeffs 1] [x 0])
  (fold-right (\l{[coeff 0] [acc 0]} (+ coeff (* x acc)))
              0
              coeffs))\end{code}

Unfortunately, using a fold operation makes it much harder to
parallelise the code.
Again, this is not particularly important if our polynomials are of low degree,
but if we were evaluating large polynomials,
we might want a function that can efficiently make use of multiple
processors to execute in parallel.
We can achieve this with our final definition:
\begin{code}
;;; Efficient on serial or parallel processor
(define (poly-eval [coeffs 1] [x 0])
  (reduce/zero + 0
               (* coeffs
                  (open-scan/zero * 1 
                                  (with-shape coeffs x)))))\end{code}
This definition uses the \ex{with-shape} function,
which uses the elements from its second argument (the scalar \ex{x})
to produce an array whose shape matches its first argument
(the \ex{coeffs} vector).
If the data array has too few elements,
\ex{with-shape} cycles through them repeatedly;
if the data array has too many elements,
then trailing, unneeded items are silently discarded.  

If our \ex{coeffs} vector has $n$ coefficients,
the \ex{with-shape} expression produces an $n$-element vector,
whose elements are all \ex{x}.\footnote{
  An alternate way to make a vector of \ex{x} values with the
  same shape as the coefficients vector would be to
  map a constant \ex{x}-producing function across \ex{coeffs},
  with the expression \ex{((\l{[y 0]} x) coeffs)}, or,
  equivalently, with the \ex{let} form of the $\beta$-redex:
\begin{code}
(let ((y 0 coeffs)) x)\end{code}
%Either expression will permit the type system to determine that
%the shape of the result vector matches the shape of the coefficients vector.
}
We then use the \ex{open-scan/zero} function to multiply together
the elements of this vector.
This scan operator is an external scan that produces a result with
the same number of elements as the input;
it does this by not producing the final element of the full scan.
The result is a vector of exponentiations \ex{[$x^0$ $x^1$ {\ldots} $x^{n-1}$]}.
Computing this with a scan has two advantages:
we only do a linear number of multiplies,
and we do so in a fully parallelised manner.
From here, the code is straightforward: we multiply the vector of $x$-powers
by the coefficients vector and sum the terms.
Note that the summation is done with a reduction,
so this part of the computation is also parallelisable.
And, of course, the data-parallel bits of the computation
expressed with basic rank-polymorphic frame/cell distribution are
trivially parallelisable, as well.
(In the case of this code, this is the point-wise multiplication
of the coefficients by the $x$-powers vector.)

%%%%%%%%%%%%%%%%%%%%%%%%%%%%%%%%%%%%%%%%%%%%%%%%%%%%%%%%%%%%%%%%%%%%%%%%%%%%%%%
\Section{Conditional code in Remora}
\label{sec:conditional-code}

Conditional code in rank polymorphic languages such as Remora is
done a bit differently,
in order to express the computation in a parallel way.
Remora provides the exact same \ex{if} and \ex{cond} forms
that \textsc{Lisp} and Scheme provide,
but these forms are considered barriers to parallelism.
So we can write a serial version of factorial as follows:
\begin{code}
(define (fact [n 0])
  (if (zero? n) 1
      (* n (fact (- n 1)))))\end{code}
The previous version of factorial that uses \ex{iota} and \ex{reduce/zero}
(page~\pageref{example:factorial})
permits the multiplications to be done in parallel.

More in line with the general data-parallel focus of the language,
we can filter a collection of items by some criterion using the
\ex{filter} function.
This code shows how to use \ex{filter}
to select out all the positive elements of a vector:
\begin{code}
    (define nums [0 5 -7 -22 91 100])

    (filter (> nums 0) nums)
\result{[5 91 100]}\end{code}

The \ex{filter} function takes a boolean ``selection'' vector and a data array;
the boolean vector must be as long as the initial dimension of the data array
(so the data array must have rank greater than zero---it
may not be a dimensionless scalar).
If the boolean vector has $d_1$ items,
and the data array has shape $[d_1, d_2,\ldots,d_r]$,
the function views the data array as a collection of $d_1$ sub-arrays,
each with shape $[d_2,\ldots,d_r]$;
it selects out all the sub-arrays whose corresponding boolean is true.

So we can select certain rows of a matrix with
\label{example:matrix-filter}
\begin{code}
    (filter [#t #f #f #t #t]
            [[ 0  1  2]    ; yes
             [16 17 18]    ; no
             [ 9 10 11]    ; no
             [22 23 24]    ; yes
             [96 97 98]])  ; yes
\result{[[ 0  1  2]
 [22 23 24]
 [96 97 98]]}\end{code}
A similar function, \ex{partition},
takes the same arguments and splits the data set into two collections,
then uses Remora's multiple-value return feature to return two
distinct arrays.

To select columns from a matrix, rather than rows, we rerank the
function so that it is individually applied to each row:
\begin{code}
    (~(1 1)filter [ #t #f #t]
                  [[ 0  1  2]
                   [16 17 18]
                   [ 9 10 11]])
\pagebreak[0]
\result{[[ 0  2]
 [16 18]
 [ 9 11]]}\end{code}

The simple function \ex{select} takes a boolean and two other values,
returning the first if the boolean is true, and the second, if it is
false.
This function is fundamentally defined on scalars,
but it is frequently employed in a lifted capacity:
\begin{code}
    (select #t 3 4)
\result{3}

    (select #f [8 1 2] [20 3 9])
\result{[20 3 9]}

  (select [#t #f #f #t #t]
          [ 0  1  2  3  4]
          [20 21 22 23 24])
\result{[0 21 22 3 4]}

  ;; Merge a default value into a data collection
  (select [#t #f #f #t #t]
          [ 0  1  2  3  4]
          100)
\result{[0 100 100 3 4]}\end{code}

A generalisation of \ex{filter} is the \ex{replicate} function.
Instead of taking a vector of booleans, it takes a vector of
counts that specify how many copies of each item to provide:
\begin{code}
    (replicate [1    3   0   2]
               [20  73  99  14])
\result{[20 73 73 73 14 14]}\end{code}

An example of using these data-parallel operators to write conditional
code is a function that takes a string (that is, a vector of characters)
and converts each contiguous run of spaces into a single space:
\begin{code}
;;; Keep a char if it's not a space, or
;;; its left neighbor isn't a space.
(define (collapse-spaces [s 1])
  (let* ((ns  (not (char=? #\\space s)))
         (lns (drop-right1 (append [#t] ns) 1))))
    (filter (or ns lns) s)))\end{code}
The \ex{drop-right1} function takes a single drop count
(hence the ``1'' in its name) which says how to shorten
the leading dimension of its initial argument.
(There is a more general \ex{drop-right} that permits the programmer
to provide a vector of drop counts, one for each dimension of the array.)
In our example above, we are shifting the \ex{ns} vector right by one element, 
filling in on the left with an initial true value (using \ex{append}),
and discarding the rightmost element (using \ex{drop-right1}).

As usual, this function is completely parallelisable
(since \ex{filter} can be implemented using similar techniques to
parallel reductions and scans);
it is defined without recourse to iteration or array indexing.

%%%%%%%%%%%%%%%%%%%%%%%%%%%%%%%%%%%%%%%%%%%%%%%%%%%%%%%%%%%%%%%%%%%%%%%%%%%%%%%
\Section{Indexing items from arrays}
\label{sec:indexing}

Remora provides the ability to index into arrays.
However, programs that use indexing to fetch a single element from an array
at a time will not be able to operate on the array in parallel,
so this is generally considered bad programming style,
counter to Remora's general data-parallel paradigm.

The proper way to use indexing is to employ it in a lifted manner:
instead of using a \emph{single} index to fetch a \emph{single} value
from an array, use an \emph{array} of indices to fetch a
\emph{collection} of values from an array.

Let's define a three-dimensional $3 \times 2 \times 4$ array
to use for an example.
\begin{code}
    (define a [[[ 1   10  100 1000]  ; plane 0
                [ 2   20  200 2000]]

               [[ 0    2    4    6]  ; plane 1
                [ 1    3    5    7]]

               [[30   31   32   33]  ; plane 2
                [40   41   42   43]]])\end{code}
We can select the element in plane 1, row 1, column 2 with
\begin{code}
    (index a [1 1 2])
\result{5}\end{code}
Note that the returned value is an array, of course---all
expressions in Remora produce array results
(in this case, a scalar array).

The \ex{index} operation is just a function,
so it lifts like any function in Remora.
If we use a vector of indices
(that is, a vector of index vectors, which is to say, a matrix),
we will get a vector of results:
\begin{code}
    (index a [[1 1 2]
              [1 1 2]
              [0 1 3]])
\result{[5 5 2000]}\end{code}

If we use a matrix of indices (that is, a rank-3 array),
we will get a matrix of results:
\begin{code}
    (index a [[[1 1 2]  [0 1 3]]
              [[2 0 0]  [1 0 3]]])
\result{[[5  2000]
 [30    6]]}\end{code}

The lifting rule for indexing is that the \ex{index} function
takes its entire first argument as its source-array cell,
and breaks its second argument up into a frame of index vectors.
However, we can rerank \ex{index} to break the first, source-array
argument up into smaller cells.
So, to fetch the element at row 0 column 2 from every plane of \ex{a},
we rerank \ex{index} to fetch from matrices:
\begin{code}
    (~[2 1]index a [0 2])
\result{[100 4 32]}\end{code}

The index function can select out an entire subarray from an array,
if we provide it with an index vector that is shorter than the full
rank of the array.
To select plane 2 of \ex{a}, we write
\begin{code}
    (index a [2])
\result{[[30 31 32 33]
 [40 41 42 43]]}\end{code}
Likewise, we can select plane 2 row 1 with
\begin{code}
    (index a [2 1])
\result{[40 41 42 43]}\end{code}
% I removed this because [] is rank-ambiguous, so you'd have to
% write the empty vector (array [0]) and I don't want to get into that.
%As a boundary case, an empty index vector 
%simply returns the entire source array.
%\begin{code}
%    (index a [])
%\result{
%[[[ 1   10  100 1000]  ; plane 0
%  [ 2   20  200 2000]]
%
% [[ 0    2    4    6]  ; plane 1
%  [ 1    3    5    7]]
%
% [[30   31   32   33]  ; plane 2
%  [40   41   42   43]]]}\end{code}

Note that the \ex{index} function considers an index to be a vector
of integers.
As a special case, the function \ex{index-item} takes a scalar integer
as index and uses it to select along the first dimension of the source
array.
So we can select plane 1 of array \ex{a} with
\begin{code}
    (index-item a 1)
\result{[[0 2 4 6]
 [1 3 5 7]]}\end{code}

\Section{Indexing and performance}

Pumping a large collection of indices through an array is a 
powerful computational act: we can rotate, mirror, replicate,
select, sort, and arbitrarily permute the elements of an array
by performing various calculations on arrays of indices and then
performing a single group index operation.
This power comes at a cost, however:
it is not likely that a compiler will be able to guarantee locality for
the sequence of element fetches required, nor will it be able
to guarantee statically that each index is a legal one, so the
individual fetches will all have to be checked for safety.

So, it is best to regard Remora's lifted \ex{index} operation as,
in fact, a big \emph{communications} step, that shuffles data
from one layout scattered across the processors executing the
program to a new layout;
such a step taxes the computer's communications hardware, 
not its computational resources.

The more structured array manipulations performed by Remora's
library of \ex{rotate}, \ex{mirror}, \ex{take}, \ex{drop}, {\etc}
functions can provide better performance for both of these reasons
(locality and elimination of index-safety checks).

\Section{Selecting regions of an array}
An operation related to indexing is the family of \ex{subarray} functions.
While \ex{index} produces a single scalar, 
these functions return an array whose rank is the same as the input array.

For example, we can select a $2 \times 2 \times 2$ cube out of array \ex{a}
starting at index \ex{[1 0 2]} with
\begin{code}
    (subarray a [1 0 2] [2 2 2])
\result{[[[4 6]
  [5 7]]

 [[32 33]
  [42 43]]]}\end{code}

The second argument to \ex{subarray} is the \emph{start} index;
the third argument is the \emph{shape} vector.
The shape vector can be shorter than the rank of the source array;
unspecified elements select the entire corresponding dimension of the
array.
Thus we can select out two planes of \ex{a} starting at index \ex{[1 0 2]}
with
\begin{code}
    (subarray a [1 0 2]] [2])
\result{[[[ 4  6]
  [ 5  7]]

 [[32 33]
  [42 43]]]}\end{code}
This gets us 2 planes beginning with plane 1;
all rows beginning with row 0; and
all columns beginning with column 2.
Note that the initial elements of the shape of the result array
are given by the shape argument (in this case, \ex{[2]}).

The variant function \ex{subarray/wrap} permits the programmer to select
subarrays with toroidal topology, or wrapping on dimensions, if the selected
subarray selects beyond the end of some dimension of the source array.

The variant function \ex{subarray/fill} permits the programmer to
specify a default ``fill'' value if the selected subarray extends
beyond the end of some dimension of the source array.

\Section{Sorting}

The \ex{grade} function sorts data;
it returns a permutation vector of integer indices
that can then be used with \ex{index-item} to shuffle the data.

\begin{code}
    (define v [3 1 4 1])
    (grade < v)
\result{[1 3 0 2]}\pagebreak[0]
    (index-item v (grade < v))
\result{[1 1 3 4]}
    (index-item v (grade > v))
\result{[4 3 1 1]}\end{code}

The \ex{grade} function provides a stable sort. 
The comparison function is applied to the ``items'' of the data array,
that is, if the data array has rank $r$, it is applied to subarrays
of rank $r-1$.

So we can sort the rows of a matrix into ascending order,
using a lexicographic sort to compare two rows, with
\begin{code}
    ;;; Lexicographic order for two vectors
    ;;; The reduction picks out the leftmost non-zero
    ;;; element of its input vector v-w.
    (define (lex< [v 1] [w 1])
      (negative? (reduce (\l{x y} (select (zero? x) y x))
                         (- v w))))

    (define m [[3 1 4 1 5]
               [2 7 1 8 3]
               [1 6 1 8 0]])

    (index-item m (grade lex< m))
\result{[[1 6 1 8 0]
 [2 7 1 8 3]
 [3 1 4 1 5]]}\end{code}

If we just want the sorted data and don't need the permutation index vector, 
we can directly use the \ex{sort} function:
\begin{code}
    (sort < [3 1 4 1])
\result{[1 1 3 4]}\end{code}

\newpage
%%%%%%%%%%%%%%%%%%%%%%%%%%%%%%%%%%%%%%%%%%%%%%%%%%%%%%%%%%%%%%%%%%%%%%%%%%%%%%%
\Section{Explicitly typed Remora}

The code examples we've seen up to this point have been written
using a variant of the language that is \emph{dynamically typed},
in the same way that Scheme, \textsc{Lisp}, or JavaScript are.
However, serious programming in Remora is done in a language that comes with
a type system---one that is expressive enough to include descriptions of
the ranks and shapes of the arrays on which functions operate.
This means that a well-typed program comes with a guarantee that a large
class of errors involving functions being applied to arrays with incompatible
shapes cannot happen at run time.

The down side of the type system's complexity and expressiveness
is that Remora's type system is quite verbose---in some cases, types can 
be larger than the terms they annotate.
Remora is intended to be programmed by humans in a variant of the
language that supports \emph{type inference:}
the programmer writes code with no (or few) type annotations,
and a powerful inference algorithm fills in the missing type machinery,
producing a fully typed term.
The design and implementation of type-inference technology for Remora
is the subject of current research~\cite{Slepak:PhD}
and beyond the scope of this tutorial.

Even in a type-inferred context, however,
there are occasions where programmers may wish to include
some explicit types in their programs
(\eg, at module boundaries), 
so the implicitly typed
variant of the language includes the ability to mix explicitly typed
and type-inferred code.
Also, when type-inferenced code has array-shape bugs,
these will be reported to the programmer as type errors
(analogously to the way that type-inferred SML or OCaml code is
handled).
So even when programmers don't write types in their source code,
types will still show up in the debugging process.

In this section, we'll introduce Remora's type system by means of a series
of examples of steadily increasing complexity.
Keep in mind, however, that while types in Remora are a powerful skeleton that
provides Remora's iteration and parallelism control structure,
these types are typically something that---like a human skeleton---are
intended to operate hidden from view.

\Section{Types and indices}

Remora's type system is a kind of (restricted) dependent type system,
parameterised in two ways: over other types, and over ``indices,''
which represent array dimensions and shapes.
\begin{itemize}
\item A \emph{dimension index} describes a single dimension of an
      array. A dimension can be a constant, literal dimension, such
      as \ex{3} or \ex{17}; an index variable, such as \ex{d1} or
      \ex{size}; or the sum of dimensions, such as \ex{(+ d1 7)}.

\item A \emph{shape index} represents a contiguous sequence of dimensions
      from an array. A shape can be given by listing a sequence of
      dimension indices in a \ex{shape} form, \eg,
      \ex{(shape 50 d2 12)}.
      We can also write an index variable, such as \ex{@s1} or \ex{@dims},
      for an entire shape,
      or append shapes together with the \ex{++} shape operator,
      \eg, \ex{(++ @s1 (shape 3 1))}.
\end{itemize}
Note that the two kinds of index variable, dimension and shape,
are or\-tho\-graph\-i\-cal\-ly distinguished:
shape-index variables begin with an ``\ex{@}'' character, while
dimension-index variables do not.

Now that we've spelled out the core syntax for indices, we'll shift to
a syntactic sugar that is more readable: ``splicing shape notation.''
In a syntactic context where a shape can occur,
a square-bracket delimited sequence of shape and dimension indices stands
for the sequence of those indices, 
with shape indices ``spliced'' into the sequence.
Figure~\ref{fig:splicing-notation} gives examples showing the
correspondence between the splicing notation and the core notation.
\begin{figure}
\begin{inset}
\begin{tabular}{l@{\quad$\Rightarrow$\quad}l}
\multicolumn{1}{l}{Splicing-shape notation} & Core notation \\ \hline
\ex{[d1 @s 5 (+ 1 d2)]} & {
\begin{codebox}[t]
(++ (shape d1)
    @s
    (shape 5 (+ 1 d2)))\end{codebox}
} \\
\ex{[3 4]}        & \ex{(shape 3 4)}    \\
\ex{[@s1 @s2]}     & \ex{(++ @s1 @s2)}  \\
\ex{[@s]}          & \ex{@s}            \\
\end{tabular}
\end{inset}
\caption{The splicing-shape notation is syntactic sugar for writing
         array shapes.}
\label{fig:splicing-notation}
\end{figure}

Remora has two kinds of types.
First, we have types that describe the individual elements of arrays,
which includes basic primitives, such as \ex{int}, \ex{char} and \ex{bool},
and type variables, such as \ex{t1} or \ex{elt}.
We can write the type of a function that takes arguments of
(array) type $\tau_1, \ldots \tau_n$ 
and produces a result of (array) type $\tau_{\text{result}}$ with the
type
\begin{code}
\ft{\vi{\tau}{1} {\ldots} \vi{\tau}{n}}{\vi{\tau}{\text{result}}}\end{code}

The basic array type is written \ex{(A \v{t} \v{shape})},
where \v{t} is the (element) type of the array's constituent atoms,
and \v{shape} is the shape of the array.
So a $3 \times 4$ array of integers has type \ex{(A int (shape 3 4))},
while a boolean scalar has type \ex{(A bool (shape))}.
Expressions can be polymorphic over array types;
like shape-index variables,
array-type variables must be written with an initial ``\ex{@}'' character,
\eg, \ex{@a1} or \ex{@x}.

Thus, the type of the addition function is
\begin{code}
\ft{(A int (shape)) (A int (shape))}{
    (A int (shape))}\end{code}
That is, the addition function takes two arguments,
each a scalar array (with empty shape \ex{(shape)}) of \ex{int} elements,
and returns another scalar array of \ex{int}.

Beyond this core syntax, we use the syntactic sugar
\begin{code}
[\(t\) \vi{i}{1} \ldots \vi{i}{n}]\end{code}
to mean the array type \ex{(A \(t\) [\vi{i}{1} \ldots \vi{i}{n}])}.
So we can write the type of the addition function a bit more succinctly as
\begin{code}
\ft{[int] [int]}{[int]}\end{code}
Finally, if we write an element type in a syntactic context where an array
type is expected, it is taken to mean a scalar array of that type.
So we can finally reduce the type of the addition function simply to
\begin{code}
\ft{int int}{int}\end{code}
However, this is all convenience syntax which should be taken to
be shorthand for the original type of the three above.

We also have types for type-polymorphic expressions:
\begin{code}
\faty{\vi{tv}{1} {\ldots} \vi{tv}{n}}{\(\tau\)}\end{code}
Applying a value of this type to $n$ types
will produce a value of type $\tau$,
with the type variables $\vi{tv}{i}$ instantiated to the given types.
Note that the type variables $\vi{tv}{i}$ can be both element-type variables,
and array-type variables.
      
Likewise, we have a type constructor for describing types that are dependent
on (static) array dimension and shape indices:
\begin{code}
\pity{\vi{iv}{1} {\ldots} \vi{iv}{n}}{\(\tau\)}\end{code}
is a type $\tau$ abstracted over the index variables $\vi{iv}{i}$,
which can be both dimension or shape indices.

\Section{Remora and ASCII}

As we move into the details of the explicitly typed language,
we'll increasingly see non-Roman, ``mathematical'' characters,
such as $\Pi$ and $\forall$
(as well as the $\lambda$ keyword we saw
in the simpler, dynamically typed language).
There are seven such keywords in the language, 
as shown in Table~\ref{table:ascii-keywords};
all seven have simple ASCII equivalents.
%%%%%%%%%%%%%%%%%%%%
\begin{table}
\begin{center}
\makebox[\textwidth]{
\hfill
\begin{oldtabular}[t]{@{}l@{\quad}l@{}}
Keyword         & ASCII         \\\hline
$\lambda$       & \ex{fn}       \\
T$\lambda$      & \ex{t-fn}     \\
I$\lambda$      & \ex{i-fn}     \\
\tu             & \ex{->}
\end{oldtabular}
\hfill
\begin{oldtabular}[t]{@{}l@{\quad}l@{}}
Keyword         & ASCII         \\\hline
$\forall$       & \ex{Forall}   \\
$\Pi$           & \ex{Pi}       \\
$\Sigma$        & \ex{Sigma}
\end{oldtabular}
\hfill
}
\end{center}
\caption{Remora has seven core forms whose identifying keywords
use non-ASCII Unicode characters; all have ASCII equivalents.}
\label{table:ascii-keywords}
\end{table}
%%%%%%%%%%%%%%%%%%%%
We'll consistently use the more attractive Unicode / super-ASCII keywords 
in this tutorial,
but it is worth noting that Remora programs can easily be written
in the least-common denominator character set of ASCII\@.

\Section{Explicitly typed functions}
Remora provides three kinds of functions---abstractions
over values, types and indices---and three corresponding forms
for applying them.
Table~\ref{table:three-abstractions} shows the syntax of all six
forms, and the type constructors for each kind of function.
\begin{table}
{\small
\begin{tabular}{l@{\quad}l@{\quad}l@{\quad}l}
Arguments       & Function form & Application form & Function type\\
\hline
Values          & \ex{(\l{[\v{v} \v{\tau}] \ldots} \v{e})}
                & \ex{(\v{f} \v{arg} \ldots)}
                & \ex{\ft{\(\tau\) \ldots}{\v{\tau}}}
                \\
Types           & \ex{(T\l{\v{tvar} \ldots} \v{e}})
                & \ex{(t-app \v{f} \(\tau\) \ldots)}
                & \ex{\faty{\v{tvar} \ldots}{\v{\tau}}}
                \\
Indices         & \ex{(I\l{\v{ivar} \ldots} \v{e}})
                & \ex{(i-app \v{f} \v{i} \ldots)}
                & \ex{\pity{\v{ivar} \ldots}{\v{\tau}}}
\end{tabular}
}
\caption{Remora has three distinct kinds of functions or abstraction,
depending on what kind of arguments are given to the function:
values, types or indices.}
\label{table:three-abstractions}
\end{table}

An example of a simple, monomorphic value abstraction is
a function that takes a scalar integer and doubles it:
\begin{code}
(\l{[i int]} (+ i i))\end{code}
which desugars (as an expression) to the scalar array
\begin{code}
(array []
  (\l{[i (A int (shape))]}
    (+ i i)))\end{code}
The function has type \ex{\ft{int}{int}}.
We can give the function a name with a
\ex{define} form
\begin{code}
(define double (\l{[i int]} (+ i i))\end{code}
which can also be written as a ``functional'' definition with
the syntactic sugar
\begin{code}
(define (double {[i int]})
  (+ i i))\end{code}
We apply a function to arguments simply by putting 
the function-producing expression and the argument-producing
expressions in a form with no keyword at all.
Thus, the expression
\begin{code}
(double [3 2])\end{code}
applies our doubling function to a vector.
Remora's lifting semantics maps the scalar-consuming doubler
across a vector frame, producing the result \ex{[6 4]}.

Another example showing the application of a value-consuming function
is the body expression of our doubler function, \ex{(+ i i)}.
The function-producing subexpression \ex{+} evaluates to a
scalar array whose single element is the addition function.
The two argument-producing expressions \ex{i} and \ex{i} each produce
a scalar integer array
(as determined by the type specified for \ex{i}).
The function is then applied to the two integer scalar arguments.

We can write functions that are polymorphic---that is, abstracted
over \emph{type} arguments instead of value arguments---with the
form \ex{(T\l{\v{tvar} \ldots}{\v{body}})}.
A very simple example is the polymorphic identity function
\begin{code}
(define identity (T\l{@t}
                   (\l{[x @t]} x))\end{code}
which has type
\begin{code}
\faty{@t}{\ft{@t}{@t}}\end{code}

We must apply this function to a type with a \ex{t-app} form before
we can apply it to an actual array:
\begin{code}
((t-app identity [bool 2 3]) [[#f #t #f]
                              [#t #t #t]])\end{code}
This code first applies \ex{identity} to an array type which
specialises \ex{@t} to boolean matrices with two rows and three
columns;
the result is an identity function 
\begin{code}
(\l{[x [bool 2 3]]} x)\end{code}
which can only be applied to $2\times 3$ matrices of booleans.
(Recall that the array type \ex{[bool 2 3]} is syntactic sugar for
the type
\ex{(A bool (shape 2 3))}.)

Note that type abstractions can have type-variable parameters
that represent element types
(written without a leading at-sign: \ex{t}, \ex{x}, \etc),
or full array types
(written with a leading at-sign: \ex{@t}, \ex{@x}, \etc).
                              
As another example, consider the polymorphic function that is a
``function doubler''---that is, it takes a function $f$ and
returns the function which applies $f$ twice to its input.
This function can be defined and given a name with
\begin{code}
(define fdouble (T\l{@t}
                  (\l{[f \ft{@t}{@t}]}
                    (\l{[x @t]} (f (f x)))))\end{code}
It has type
\begin{code}
\faty{@t}{\ft{\ft{@t}{@t}}{
            \ft{@t}{@t}}}\end{code}

As with our identity function, using \ex{fdouble} requires a double
application.
First, we use a \ex{t-app} to specialise the polymorphic function
to a given type (such as floating-point scalars),
then we apply the resulting function to the function we wish to double
(in the following example, the sine function):
\begin{code}
(define sine-twice ((t-app fdouble [float]) sin))

(sine-twice [[3.14159 2.71828]
             [0.0     1.0    ]])\end{code}
Because \ex{sine-twice} is typed to operate on scalar arrays,
we are allowed to apply it to a matrix of floating-point elements
by means of Remora's rank-polymorphic function-call machinery---the
$2\times 2$ array is treated as a matrix frame of scalar cells,
so we apply the sine function twice to each of the cells.

Finally, Remora permits the programmer to write functions that
abstract over shape and dimension indices.
%
%Remora has three different function-application forms,
%one for applying functions with $\forall$ type to type arguments
%\begin{code}
%(t-app \v{f} \vi{targ}{1} \ldots)\end{code}
%one for applying functions with $\Pi$ type to index arguments
%\begin{code}
%(i-app \v{f} \vi{\i}{1} \ldots)\end{code}
%and the default, keyword-less form for applying functions with $\tu$ type
%to values
%\begin{code}
%(\v{f} \vi{arg}{1} \ldots)\end{code}
%
Our \ex{dot-product} function is an example of a dimension-polymorphic
function: it operates on two arguments of fixed rank
(they are both rank-1 vectors).
The two arguments can be of any length, but must both be of the same
length.
Therefore, the type of \ex{dot-product} is parameterised over
the dimension index \ex{len}, which represents the length of the two vectors
\begin{code}
\pity{len}{
  \ft{[int len] [int len]}{
      int}}\end{code}
This means that when we use \ex{dot-product},
we must first apply it (using the \ex{i-app} index-application form)
to a dimension, and then apply the resulting value to the actual
vectors:
\begin{code}
((i-app dot-product 3) ; Specialise function for 3-vectors, 
 [8 1 2]               ; then apply to
 [2 0 9])              ; two such vectors.\end{code}
In this example, the \ex{3} is not the \emph{value} three---%
something we could multiply times four, or square to make nine.
It occurs in the grammatical context of an \emph{index},
so it denotes the \emph{array dimension} three.

As another example, consider the \ex{append} function,
which appends two arrays of identical rank 
along their principal, or initial, dimension.
For example, if we append a $2 \times 3 \times 5$ array of booleans
and a $7 \times 3 \times 5$ array of booleans,
we get a result which is $9 \times 3 \times 5$.
So, the type of the \ex{append} function is
\begin{code}
\pity{da db @rest}{
  \faty{t}{
    \ft{[t da @rest] [t db @rest]}{
        [t (+ da db) @rest]}}} \end{code}
In our example above,
\begin{itemize}
\item index \ex{da} would be $2$, the initial dimension of the first argument,
\item index \ex{db} would be $7$, the initial dimension of the second argument,
\item index \ex{@rest} would be $[3, 5]$, the remaining shape of both arguments, and
\item type \ex{t} would be \ex{bool}, the type of the arrays' elements.
\end{itemize}
Here is an expression that uses \ex{append} to concatenate two arrays
\ex{a1} and \ex{a2} of the given shapes:
\begin{code}
((t-app (i-app append 2 7 [3 5])
        bool)
 a1 a2)\end{code}
This triple index/type/value application is
a pattern that frequently occurs when using shape-polymorphic functions.

\Section{The types of scan and reduce functions}

The scan / reduce / fold / trace family of functions are
all both type- and index-polymorphic.
In the terminology of these functions, 
we think of the source array as a vector of subarrays,
called the ``items'' of the reduction,
to be combined by the reduction operator.
Thus, if we want to add together the planes of a 3-D array,
the items of the array are its individual planes,
each a 2-D matrix.

As a convenience, 
the types for these functions include an extra shape-index parameter
that permits us to ``pad'' a ``sub-item'' operator like scalar \ex{+} 
when we want to use it to combine together higher-rank items.
The full type for \ex{reduce} is
\begin{code}
\pity{d-1 @item-pad @cell-shape}{
  \faty{t}{
    \ft{ \ft{ [t @cell-shape] [t @cell-shape] }{
              [t @cell-shape]}
          [t (+ d-1 1) @item-pad @cell-shape] }{
        [t @item-pad @cell-shape])}}}\end{code}

Suppose, for example, that we wish to sum the planes of 
a 3-D array \ex{A} with shape \ex{[5 7 4]}.
We want to use the scalar \ex{+} directly,
rather than have to ``up-rank'' the operator to work on 
$7 \times 4$ matrices by tediously writing out the $\eta$-expanded
\begin{code}
(\l{[x [int 7 4]] [y [int 7 4]]}
  (+ x y))\end{code}
This is the purpose of the \ex{@item-pad} shape in the type of
\ex{reduce}: 
the operator we use to reduce the input array can operate
on sub-items of the array, of shape \ex{@cell-shape};
it will be padded up to the full items of the array with the
extra dimensions of \ex{@item-pad}.

So, in our $5 \times 7 \times 4$ array example, 
we have $\ex{d-1} = 4$, 
which means the initial or ``iteration'' dimension
along which we reduce is \ex{(+ d-1 1)}, or $5$,
as is specified on line 4 of the type.
Adding one in the type is how we require this dimension of the array
to be at least one: 
\ex{reduce} cannot operate on arrays where 
the iteration dimension is zero.
(The intent of choosing ``\ex{d-1}'' for the name of the index variable 
is that it be read as ``d minus 1.'')

Continuing through the parameters of our type,
the shape index \ex{@cell-shape} provides the shape of the
arguments consumed and produced by the reduction's combining operator.
Since our combining operator \ex{+} works on scalars,
we have $\ex{@cell-shape} = \ex{[]}$ and type $\ex{t} = \ex{int}$.
However, we will actually be adding together $7 \times 4$ matrices,
not scalars, so we must pad the shape up by specifying 
$\ex{@item-pad} = \ex{[7 4]}$.

Thus, to sum the planes of our $5 \times 7 \times 4$ array \ex{A}, 
we write
\begin{code}
((t-app (i-app reduce 4 [7 4] []) int) + A)\end{code}

The types of the other reduction, scan, fold and trace operators are
similar in structure to the type of \ex{reduce},
\emph{mutatis mutandis.}
%%%%%%%%%%%%%%%%%%%%%%%%%%%%%%%%%%%%%%%%%%%%%%%%%%%%%%%%%%%%%%%%%%%%%%%%%%%%%%%
\Section{Boxes and existential types}
\label{sec:boxes}

``Boxes'' are a mechanism in Remora that permit the language
to handle two issues:

\begin{itemize}
\itum{Array dimensions are not always statically known}
Remora's index language gives programmers a limited ability to compute
the shapes of arrays we are calculating from the shapes of the arrays
that form the inputs to our calculations,
using the \ex{+} operator to add dimensions,
and the \ex{++} operator to append shapes.

Sometimes, however, the shape of some computed array is too complex
to be described by such a restricted facility.
Consider the function
(written in the dynamically typed variant of the language):
\begin{code}
(\l{[n 0]} (iota [n]))\end{code}
Trying to come up with a typed version of this function runs into a problem:
the length of the vector produced by the function is dependent on
a \emph{computed value}, not some other array shape.

Reasoning about the possible values that \ex{n} might have is
equivalent in decidability to the halting problem,
so we have no hope of being able to solve this in general.
Furthermore, Remora permits I/O, 
so the value of \ex{n} could be supplied at run time
by a human sitting at a keyboard,
thus rendering it completely unknowable at compile time.

\itum{Some data collections are not regular arrays}
In Remora, we can write a vector of characters, \eg, the five vowels
\begin{code}
[#\\a #\\e #\\i #\\o #\\u]\end{code}
more compactly as \ex{"aeiou"}.
So \ex{["abc" "xyz"]} is a matrix with two rows and three columns.

Now, suppose we wish to represent the five days of the work week as
character data.
We cannot write this as a matrix of characters
\begin{code}
["Monday"       ; *Not* a legal
 "Tuesday"      ; expression!
 "Wednesday"
 "Thursday"
 "Friday"]\end{code}
This is not a legal array---the third row (Wednesday) has nine columns,
while the first row (Monday) has only six.
We could consider padding short rows with some kind of null or blank filler,
but this is not a general solution to the problem of ``ragged'' data.

\end{itemize}

In Remora, a ``box'' is a thing that packages up an entire array as a single 
\emph{element} in some other array.
It is a mechanism that lets us regard a \emph{collection} (an array)
as a member (an element) of some other collection.
Boxes package up two things together:
a \emph{value} (that is, an array),
and a set of \emph{indices} that we will need later
when we \emph{unbox} the array.

We can write our days-of-the-week example using boxes as
\begin{code}
(boxes (len) [char len] [5]
  ((6) "Monday"   )
  ((7) "Tuesday"  )
  ((9) "Wednesday")
  ((8) "Thursday" )
  ((6) "Friday"   ))\end{code}
Let's begin with the type of this expression, which is
\begin{code}
[(\boxtype{len} [char len]) 5]\end{code}
This is the type of vectors of length five whose elements
each have item type
\begin{code}
(\boxtype{len} [char len])\end{code}
In English, this element type means,
``There is some dimension \ex{len} such that this box contains
  an array of type \ex{[char len]}---a character vector of length
  \ex{len}.''
What's more,
each box in our vector of five boxes can use a \emph{different dimension} for
\ex{len}.
In one box, \ex{len} can be five; in another, it can be twenty-seven; and so on.

The \ex{boxes} form shows how we can build our example boxed array:
\begin{itemize}
\item The first part of the \ex{boxes} form is a list of index variables
over which the boxed-up arrays are existentially abstracted.
In our case, we are only abstracting over the length of each boxed-up
vector, so we only have the dimension index \ex{len}.

\item The second part of the form, \ex{[char len]}, is the abstracted type:
each boxed-up array must be a character vector of length
\ex{len} for some value of \ex{len}.

\item The third part of the form, \ex{[5]}, is the shape of the array
we are constructing: it is a five-element vector.

\item Finally, we have a series of index/expression clauses.
Each clause specifies a particular dimension for \ex{len}
and an expression that has type \ex{[char len]}
\emph{given that particular \ex{len}}.
Thus, the specific dimension is called the ``witness'' for the
element type \ex{(\boxtype{len} [char len])}.
\end{itemize}

In our example, \ex{"Monday"} is a character vector of length six,
so we specify $\ex{len} = 6$ in the first clause;
likewise, we specify $\ex{len} = 8$ in the \ex{"Thursday"} clause;
and so forth.

The general syntax of such an array expression is
\begin{code}
(boxes (\v{ivar} \ldots) \(\tau\) [\v{dim} \ldots]
  ((\v{index} \ldots) \v{exp}) \ldots)\end{code}
The \ex{(\v{ivar} \ldots)} part of the form is a list of index
variables over which each box's array type is existentially
abstracted---in our example, we only needed one such index variable.
The $\tau$ is an array type which gives the type of the array
inside each box, which may refer to the \v{ivar} index vars.
The shape \ex{[\v{dim} \ldots]} is the shape of the box
array we are constructing.
The rest of the form is a sequence of indices/expression pairs
\ex{((\v{index} \ldots) \v{exp})}.
The form must have one of these pairs for each element of the array,
just as with an \ex{array} form.
Each expression $\v{exp}$ must produce an array of type $\tau$
once the witness indices are plugged into the type for the
corresponding \v{ivar} index variables.
  
As another example, here is a three-element vector of boxes,
where each element is a boxed integer matrix:
\begin{code}
(boxes (r c) [int r c] [3]
  ((2 2) [[1 2]
          [3 4]])

  ((1 3) [[10 100 1000]])
  
  ((3 1) [[10] [100] [1000]]))\end{code}

The first element of this vector is a $2 \times 2$ matrix;
the second, a $1 \times 3$ matrix;
and the third, a $3 \times 1$ matrix.

Boxes are frequently created one at a time, from some given array,
so the syntactic sugar
\ex{(box ((\v{ivar} \v{index}) \ldots) \(\tau\) \v{exp})}
is a convenient shorthand for such a scalar box array;
it is equivalent to
\begin{code}
(boxes (\v{ivar} \ldots) \(\tau\) []  ; The "[]" means a scalar array;
  ((\v{index} \ldots) \v{exp}))    ; here is its one item.\end{code}
For example, the expression
\begin{code}
(box ((len 3)) [int len] [8 23 0])\end{code}
boxes up a three-element integer vector, hiding the length of the
vector, as the single item of a scalar array of boxes.

We can unpack an array of boxes with the \ex{unbox} form.
For example, suppose our days-of-the-week vector above is named \ex{weekdays},
and we wish to compute a vector of booleans specifying which weekdays
are exactly six letters long.
We write
\begin{code}
    (unbox weekdays (day len)
      (= 6 ((t-app (i-app length len []) char) day)))

\result{[#t #f #f #f #t]}\end{code}

For each box in the vector \ex{weekdays},
the \ex{unbox} form opens up the box;
extracts the array and binds it to the variable \ex{day};
extracts the corresponding dimension witness---which, in this
example, is the length of the character vector \ex{day}---and
binds it to index variable \ex{len};
and  finally evaluates the body of the form.
Each time the \ex{=} numeric comparison is evaluated,
it produces a boolean scalar array;
these arrays are collected into the vector frame given by \ex{weekdays},
and we get a five-element vector of booleans,
informing us that only ``Monday'' and ``Friday'' have exactly six characters.

This explicitly typed code is somewhat ugly because
the \ex{length} function is type and shape polymorphic.
We have to instantiate it with an \ex{i-app} / \ex{t-app} pair
before applying to its actual value argument \ex{day}.
The full type of \ex{length} is
\begin{code}
\pity{d1 @s}{\faty{t}{\ft{[t d1 @s]}{int}}}\end{code}
(Note that \ex{length} requires its argument to have rank
greater than zero---that is the point of splitting out the initial
dimension \ex{d1} of its argument.)
The type-inferred version is easier to read:
\begin{code}
(unbox weekdays (day len)
  (= 6 (length char) day))\end{code}

Note that we cannot pass \ex{len} as a parameter to \ex{=} 
because it is an index variable, not a run-time integer value.
(Effectively, it is the business of \ex{length} to shift the array's
length from the index world down to the value world.)

\Section{Boxes and arrays with dynamic shape}

When we program in Remora,
array shapes are statically known much of the time.
For example, if we use \ex{reduce} to collapse vertically an array with
type \ex{[char 3 5]}
(that is, a $3 \times 5$ matrix of characters),
then we know the result will have type \ex{[char 3]}.

Sometimes, however, we need to compute an array whose shape is
determined by some complex computation that cannot be expressed
in terms of the shapes of the computation's input values using
the restricted notation of the Remora type system
(which only permits the programmer to append shapes and add
dimensions).
Other times, an array's shape may depend on values that are
supplied only at run time,
perhaps as input from the user running the program.

As described earlier, a standard example of this is \ex{iota}.
Because this function produces an array whose shape
cannot be determined at compile time, it is defined
to return this array as a scalar box,
packaged up with its shape.
Thus, the type of \ex{iota} is
\begin{code}
\pity{r}{\ft{[int r]}{(\boxtype{@s} [int @s])}}\end{code}
In English, this means that \ex{iota} takes as its argument an integer
vector of length \ex{r} which specifies
(as a run-time value, \emph{not} as a compile-time shape index)
the shape of the array to produce,
and returns a scalar box containing an integer array of some shape
\ex{@s}.
In fact, the shape \ex{@s} of this array is the same as the
input argument{\ldots} except that the former is a shape index
in our type system, and the latter is an actual value.

Note that when we open up a box made by \ex{iota} with an \ex{unbox} form
binding \ex{a} to the boxed array and \ex{@s} to its shape
\begin{code}
(unbox ((i-app iota 2) [r c])
       (a @s)
  {\ldots} a {\ldots} @s {\ldots})\end{code}
we know absolutely nothing, at the type-system level,
about the shape of \ex{a}.
Even though a human can tell from the above code that \ex{a} is
a matrix with \ex{r} rows and \ex{c} columns,
all that is known at the type/index level is that \ex{a}
\emph{has} a shape (named \ex{@s}), which is not very informative.

Remora provides a small family of restricted variants to the \ex{iota}
function, specialised to different ranks, which allow the type system
to extract more information about the shape of the result: \ex{iota0} through
\ex{iota9}.
For example, \ex{iota1} produces a (boxed) vector,
while \ex{iota2} produces a (boxed) matrix.
Thus, the type of \ex{iota1} is
\begin{code}
\ft{int}{
    (\boxtype{len} [int len])}\end{code}
This function produces a scalar box that packages up an integer
vector.
%Likewise, the type of \ex{iota2} is
%\begin{code}
%\ft{int int}{(\boxtype{r c} [int r c])}\end{code}

We can use \ex{iota1} to write the factorial function:
\begin{code}
(define (fact [n int])
  (unbox (iota1 n) (factors-1 len)
    ;; RED is type & index-specialised REDUCE/ZERO:
    (let ((red (t-app (i-app reduce/zero len [] []) int)))
      (red * 1 (+ 1 factors-1)))))\end{code}
The purpose of the \ex{let}-binding is simply to break up
the code for clarity; it is not strictly necessary.

\Section{Avoiding \ex{iota} and rank-monomorphic programming}
The iota function is frequently used in (dynamically typed) APL and J
to generate arrays that encode ``iteration space,''
as collections whose items specify locations in a sequence.
When possible, programmers in Typed Remora are better off generating
such arrays using the \ex{indices-of} function,
as the shape of the result can be determined by the type system,
at compile time, and no boxing is necessary.
See the one-dimensional convolution code on page~\pageref{example:convolve1}
for an example of \ex{indices-of} being used in this way.

Some functions in Remora that manipulate array shapes produce results
whose shapes are difficult to express in terms of the shapes of their
input arrays, given the bounded expressiveness of Remora's type
system.
One example is \ex{indices-of}.
The final \emph{dimension} of its result array is 
the \emph{rank} of its input array:
given a $7 \times 10$ matrix,
it will produce a $7 \times 10 \times 2$ result;
given a rank-3 array with dimensions $11 \times 5 \times 14$,
it will produce an $11 \times 5 \times 14$ array of 3-vectors,
that is,
a rank-4 array with dimensions $11 \times 5 \times 14 \times 3$.
Thus the type of \ex{indices-of} is
\begin{code}
\pity{@s}{
  \faty{t}{
    \ft{[t @s]}{
        (\boxtype{r}{          ; r = (length @s)
          [int @s r]})}}}\end{code}
We have to use a box type here because the type system is not
expressive enough to say that the final dimension \ex{r} of the
result array is the length of the input array's shape \ex{@s}
(that is, dimension \ex{r} is the rank of the input array).
The fact that \ex{r} is the length of shape \ex{@s} is, alas,
not expressed in the types, but rather hidden away in a comment.

True rank-polymorphic programming is rare in Remora:
programmers almost always know the ranks of their arrays.
Because of this, we can work around the limits of the type system
by providing rank-monomorphic variants of functions such as
\ex{indices-of}.
Thus, we have the nine functions \ex{indices-of/1} through
\ex{indices-of/9}, each specialised to inputs of the specified rank.
For example, \ex{indices-of/2} is restricted to operate on matrices
only; it has type
\begin{code}
\pity{d1 d2}{
  \faty{t}{
    \ft{[t d1 d2]}{   ; input d1 x d2 matrix
        [int d1 d2 2]}}}\end{code}
Note that this function, happily, does not produce a boxed result;
the result array's shape is completely known given the shape of the
input.

If we use a rank-monomorphic version of \ex{indices-of},
the fully typed version of the one-dimensional convolution given earlier is
\begin{code}
(define vector-convolve
  (I\l{vlen wlen}
    (\l{[v [int vlen]] [w [int wlen]]}
      ;; Specialise REDUCE & INDICES-OF/1
      ;; to our specific needs:
      (let ((red  (t-app (i-app reduce wlen [vlen] [])
                         int))
            (io/1 (t-app (i-app indices-of/1 wlen)
                         int)))
        ;; The actual convolution:
        (red + (* (rotate v (io/1 w)) w))))))\end{code}

Types in a Remora program constitute an ``explanation'' (or proof) to the
compiler of the shapes of the values being computed by the program.
Programming in Typed Remora requires programmers to develop a
sense of how to expose and exploit this shape information.

\Section{Conditional code and boxes}

In Typed Remora, functions such as \ex{filter} and \ex{replicate}
(see page~\pageref{sec:conditional-code})
necessarily produce boxed results,
as there is no way at compile time to determine the size of the result.
Thus, the type of \ex{filter} is
\begin{code}
\pity{da @s}{
  \faty{t}{
    \ft{ [bool da] [t da @s] }{
        (\boxtype{db} [t db @s])}}}\end{code}
That is, \ex{filter} takes a boolean vector of length \ex{da},
and an array whose initial dimension is \ex{da} and whose other
dimensions are represented by the shape \ex{@s}.
We think of this array as a collection of \ex{da} items,
each of which is an array of shape \ex{@s}.

The function uses the boolean vector to filter the collection;
we retain each subarray whose corresponding boolean is the true value.
These items are collected into an array of type \ex{[t db @s]}, 
where \ex{db} is the number of items that survived the filtering
process.
Since there is no way to determine the dimension \ex{db} at compile time,
it is hidden inside a scalar box.

If we repeat our previous example
(page~\pageref{example:matrix-filter})
filtering the rows of a $5 \times 3$ matrix, in a typed setting,
we get the following code.
The filter operation views the matrix
as a collection of five subarrays ($\ex{da} = 5$),
where each subarray is a three-element vector ($\ex{@s} = \ex{[3]}$),
whose elements are integers ($\ex{t} = \ex{int}$).
Three of the five rows are selected by the filtering operation,
so $\ex{db} = \ex{3}$;
as this cannot be determined at compile time,
the resulting collection is packaged up in a box whose type abstracts away
the dimension \ex{db}.
All the existential $\Sigma$ type tells us about the result is
that there is \emph{some} dimension \ex{db} giving the number of
rows in the $\ex{db} \times 3$ matrix.
\begin{code}
    ((t-app (i-app filter 5 [3]) int)
     [#t #f #f #t #t]
     [[ 0  1  2]    ; yes      
      [16 17 18]    ; no
      [ 9 10 11]    ; no
      [22 23 24]    ; yes
      [96 97 98]])  ; yes

\result{(box ((db 3)) [int db 3]
  [[ 0  1  2]
   [22 23 24]
   [96 97 98]])}\end{code}

\Bibliography{refs}

% LocalWords:  Remora APL Iverson hoc polymorphism dimensionality ccc
% LocalWords:  notationally subform eval vmag pointwise MIMD
% LocalWords:  iscan desireable raison etre characterises
% LocalWords:  Fortran serialised utilisation xs ys po
% LocalWords:  convolve llllll ly mor phic coeffs expt Horner's coeff
% LocalWords:  acc parallelise redex cond nums boolean generalisation
% LocalWords:  ns lns subarray toroidal lex SML OCaml parameterised
% LocalWords:  bool elt tv fn Forall arg tvar ivar monomorphic len da
% LocalWords:  subexpression fdouble specialise mutatis
% LocalWords:  mutandis decidability aeiou abc xyz vlen wlen io

\end{document}